\documentclass[floatfix,showpacs]{revtex4}
\usepackage{epsf}
\usepackage{epsfig, latexsym}
\usepackage{pictex}
\usepackage{amsmath}
\usepackage{amssymb}
\usepackage{amsfonts}
\usepackage[latin1]{inputenc}

\usepackage{graphicx}

\newcommand{\insertplot}[5]{\begin{figure}
 \hfill\hbox to 0.05in{\vbox to #5in{\vfill
 \inputplot{#1}{#4}{#5}}\hfill}
 \hfill\vspace{-.1in}
 \caption{#2}\label{#3}
 \end{figure}}
 \newcommand{\inputplot}[3]{
 \special{ps: plotfile #1}
}
\newcounter{fig}   

\newcommand{\vphi}{\varphi}

\renewcommand{\a}{\alpha}
\renewcommand{\b}{\beta}
\renewcommand{\c}{\gamma}
\renewcommand{\d}{\delta}
\newcommand{\f}{\phi}
\newcommand{\e}{\mu}
\newcommand{\g}{\nu}
\renewcommand{\l}{\lambda}
\renewcommand{\t}{\theta}

\begin{document}

\title{
Rotating Boson Stars and $Q$-Balls II:\\
Negative Parity and Ergoregions
}
\author{Burkhard Kleihaus, Jutta Kunz,}
\affiliation{ Institut f\"ur Physik, Universit\"at Oldenburg, 
D-26111 Oldenburg, Germany}
\author{Meike List, Isabell Schaffer}
\affiliation{ZARM, Universit\"at Bremen,
Am Fallturm, D-28359 Bremen, Germany}

\vspace{0.3cm}


\begin{abstract}
We construct axially symmetric, rotating boson stars
with positive and negative parity.
Their flat space limits represent spinning $Q$-balls.
$Q$-balls and boson stars
exist only in a limited frequency range.
The coupling to gravity gives rise to a spiral-like
frequency dependence of the mass and charge of boson stars.
We analyze the properties of these solutions.
In particular, we discuss the presence of ergoregions in boson stars,
and determine their domains of existence.
\end{abstract}

\maketitle

\section{Introduction}

$Q$-balls represent stationary
localized solutions in flat space.
They arise, when a complex scalar field
has a suitable self-interaction
\cite{lee-s,coleman}.
The global phase invariance of the scalar field theory
is associated with a conserved charge $Q$, 
corresponding for instance to particle number \cite{lee-s}. 

The time-dependence of the $Q$-ball solutions
resides in the phase of the scalar field
and is associated with a frequency $\omega_s$.
$Q$-balls exist only in a certain frequency range, 
$\omega_{\rm min} < \omega_s < \omega_{\rm max}$,
determined by the properties of the potential
\cite{lee-s,coleman,lee-rev,volkov,list}.
At a critical value of the frequency,
both mass and charge of the $Q$-balls assume their minimal value,
from where they rise monotonically towards both limiting values
of the frequency.
Considering the mass of the $Q$-balls as a function of the charge,
there are thus two branches of $Q$-balls, merging and ending
at the minimal charge and mass. 

The simplest type of $Q$-balls is spherically symmetric.
These possess finite mass and charge, but carry no angular momentum.
Besides the fundamental spherically symmetric $Q$-balls 
there are also radially excited $Q$-balls \cite{volkov}.
The fundamental spherically symmetric $Q$-balls 
are stable along their lower mass branch,
as long as their mass is smaller than the mass of $Q$ free bosons
\cite{lee-s}.

Recently, the existence of rotating $Q$-balls was demonstrated 
\cite{volkov,list}.  
The rotation is achieved by including an additional dependence
of the phase factor of the scalar field
on the azimuthal angle $\varphi$,
where the proportionality constant $n$ must be integer.
The resulting stationary localized $Q$-ball solutions
then possess finite mass and finite angular momentum.
Interestingly, their angular momentum $J$ is quantized
in terms of their charge $Q$,
$J=nQ$ \cite{volkov,schunck}. 
Rotating $Q$-balls with $n=1$ thus have the smallest angular momentum $J$
for a given charge $Q$.
The energy density and angular momentum density of 
rotating $Q$-balls possess axial symmetry.
There are no (infinitesimally) slowly rotating $Q$-balls \cite{volkov}.

The field equations allow for $Q$-balls 
with positive and negative parity.
The scalar field is then symmetric resp.~antisymmetric
w.r.t.~reflections \cite{volkov}.
For a given charge $Q$ one finds thus two sequences of
rotating $Q$-balls, whose angular momentum increases with $n$:
the positive parity sequence $n^+$ and the negative
parity sequence $n^-$.
While the energy density of rotating $Q$-balls with positive
parity corresponds to a torus, the energy density of rotating $Q$-balls
with negative parity corresponds to a double torus.


When the scalar field is coupled to gravity, boson stars arise 
\cite{lee-bs,lee-rev,jetzer,ms-review}.
The presence of gravity has a crucial influence 
on the domain of existence of the classical solutions.
Stationary spherically symmetric boson stars ($0^+$)
also exist only in a limited frequency range,
${\omega}_{0}(\kappa) < \omega_s < \omega_{\rm max}$,
where $\kappa$ denotes the strength of the gravitational coupling.
They show, however, a rather different type of frequency dependence.
In particular, in the lower frequency range, the boson star
solutions are not uniquely determined by the frequency.
Instead a spiral-like frequency dependence of the charge and the mass
is observed, where the charge and mass approach finite limiting values
at the center of the spiral.
Furthermore, 
the charge and the mass of these boson stars tend to zero,
when the maximal value of the frequency is approached
\cite{lee-bs,lee-rev}.

Rotating boson stars with positive parity have been
obtained before \cite{schunck,schunck2,schunck3,japan,list}.
A study of the frequency dependence
of the $1^+$ boson stars showed,
that their frequency dependence is analogous to the one of
the non-rotating $0^+$ boson stars \cite{list}.
In particular,
their charge and mass also exhibit a spiral structure.
Like rotating $Q$-balls,
rotating boson stars exhibit the angular momentum quantization,
$J=nQ$ \cite{schunck}.

Here we reconsider 
rotating boson stars with positive parity and, for the first time,
present rotating boson stars with negative parity.
We construct families of boson stars of both parities numerically
for constant values of the gravitational coupling strength
and $n=1$ and 2.
We investigate the properties of these solutions.
In particular, we analyze their frequency dependence,
and we demonstrate the appearence and development of ergoregions 
for these families of boson star solutions \cite{Cardoso}.

In section II we recall the action, the general equations of motion
and the global charges.
In section III we present the stationary axially symmetric Ansatz 
for the metric and the scalar field, we evaluate
the global charges within this Ansatz, and present the
boundary conditions for the metric and scalar field function.
We discuss $Q$-ball solutions and their properties in section IV 
and boson star solutions and their properties in section V.
Section VI gives our conclusions.

\section{Action, Equations and Global Charges}\label{c1}

\subsection{Action}\label{c1s1}

\begin{figure}[h!]
{\centerline{
\mbox{
\includegraphics[width=70mm,angle=0,keepaspectratio]{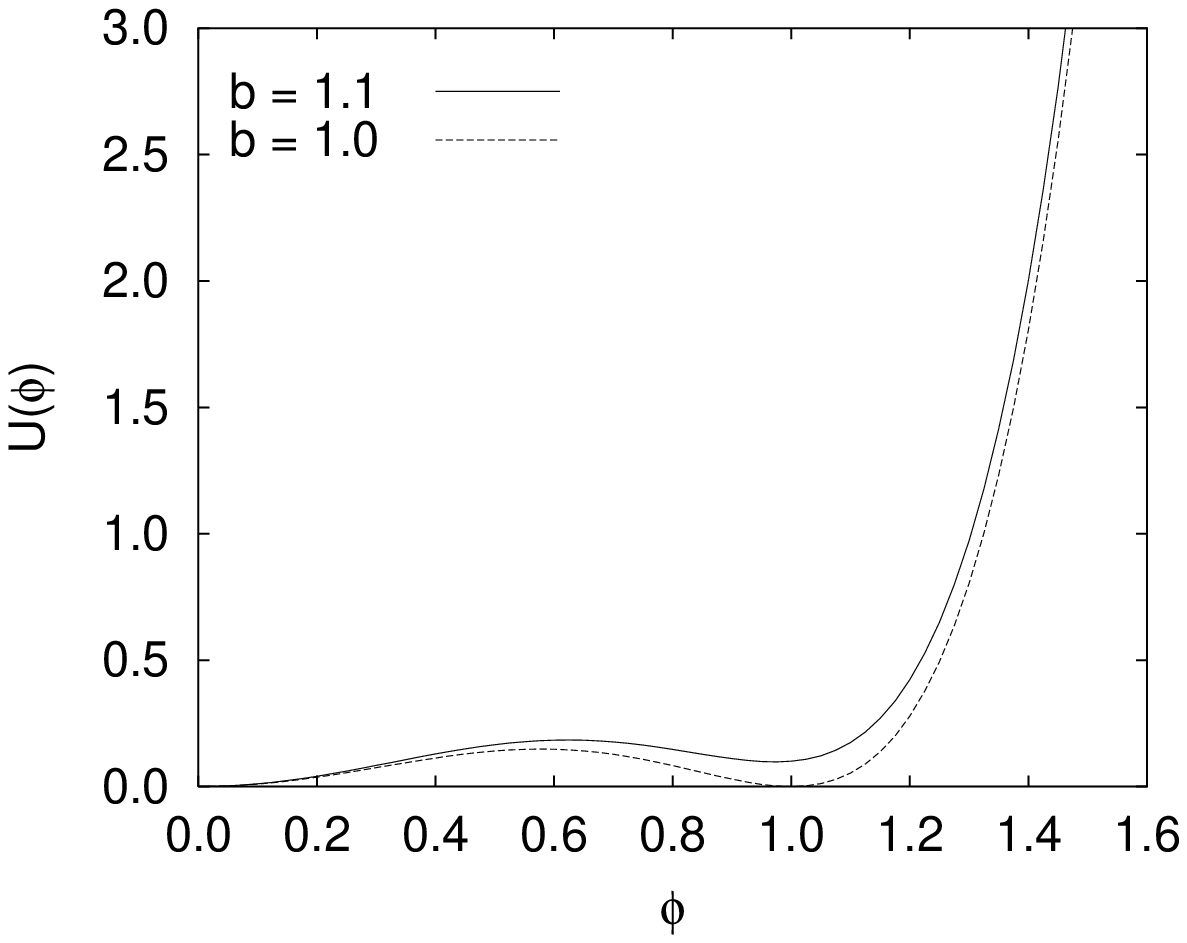}
\includegraphics[width=70mm,angle=0,keepaspectratio]{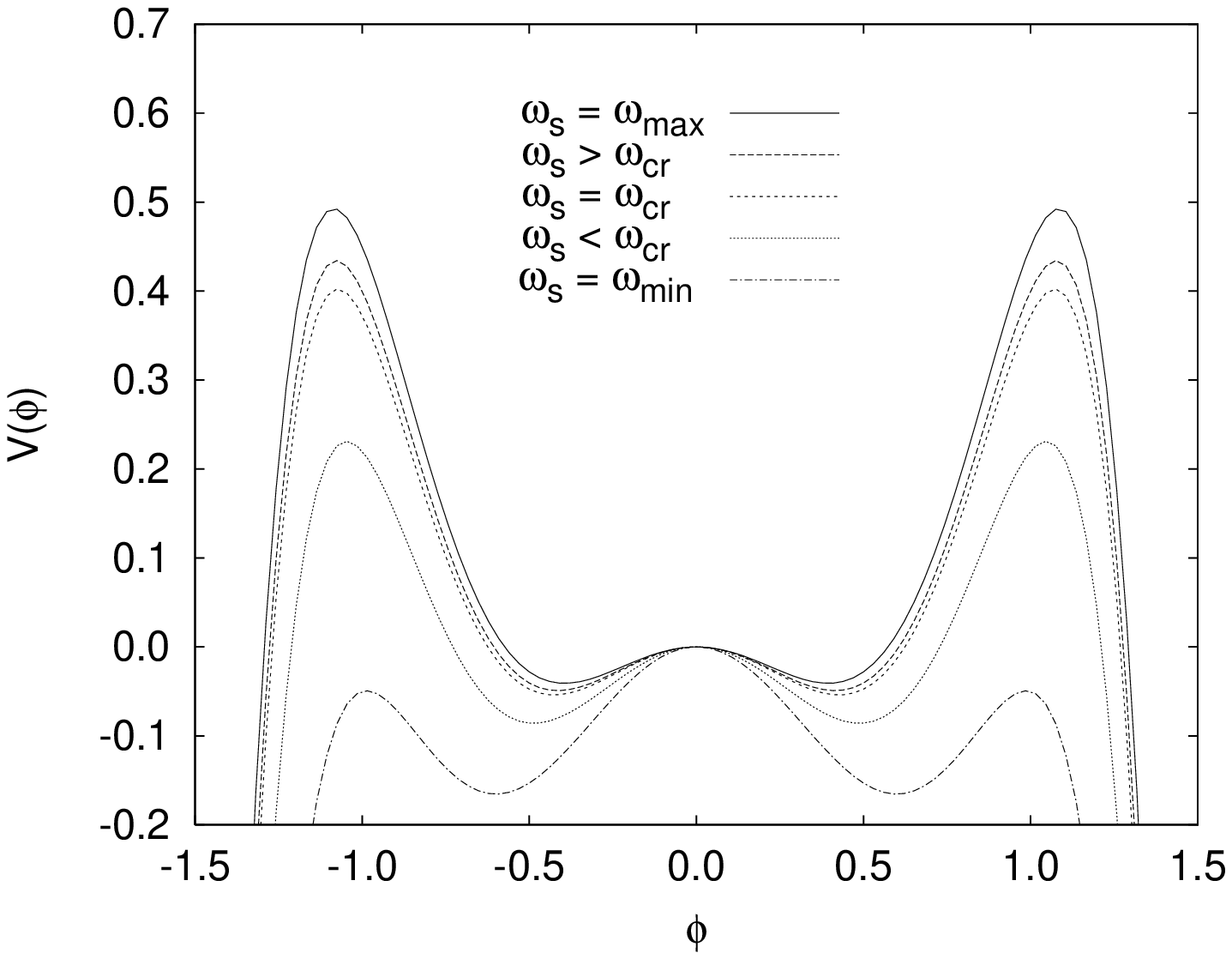}
}}}
\caption{Left: The potential $U(\f)$ versus $\f$
for $\lambda=1$, $a=2$ and $b=1.1$ resp.~$b=1$.
Right: The effective potential
$V(\phi) = \frac{1}{2} \, \omega_s^2 \, \phi^2
 - \frac{1}{2} U(\phi)$
versus $\f$
for several values of the frequency $\omega_s$.}
\label{potentiale}
\end{figure}

We consider the action of a self-interacting complex scalar field 
$\Phi$ coupled to Einstein gravity
\begin{equation}
S=\int \left[ \frac{R}{16\pi G}
   -\frac{1}{2} g^{\mu\nu}\left( \Phi_{, \, \mu}^* \Phi_{, \, \nu} + \Phi _
{, \, \nu}^* \Phi _{, \, \mu} \right) - U( \left| \Phi \right|) 
 \right] \sqrt{-g} d^4x
\ , \label{action}
\end{equation}
where $R$ is the curvature scalar,
$G$ is Newton's constant,
the asterisk denotes complex conjugation,
\begin{equation}
\Phi_{,\, {\mu}}  = \frac{\partial \Phi}{ \partial x^{\mu}}
 \ ,
\end{equation}
and $U$ denotes the potential
\begin{equation}
U(|\Phi|) =  \l |\Phi|^2 \left( |\Phi|^4 -a |\Phi|^2 +b \right)
=\lambda \left( \f^6- a\f^4 + b \f^2 \right) \ ,
\label{U} \end{equation}
with $|\Phi|=\f$.
The potential is chosen such that nontopological soliton solutions
\cite{lee-s}, also referred to as $Q$-balls \cite{coleman},
exist in the absence of gravity.
As seen in Fig.~\ref{potentiale}, 
the self-interaction of the scalar field has an attractive component,
and the potential has a minimum, $U(0)=0$, at $\Phi =0$
and a second minimum at some finite value of $|\Phi|$.
The boson mass is thus given by $m_{\rm B}=\sqrt{\lambda b}$.
In the numerical calculations of the solutions presented,
we choose the potential parameters
\cite{volkov,list}
\begin{equation}
\lambda = 1 \ , \ \ \ a=2 \ , \ \ \ b=1.1 \ .
\label{param}
\end{equation}

\subsection{Equations}\label{c1s2}

Variation of the action with respect to the metric
leads to the Einstein equations
\begin{equation}
G_{\mu\nu}= R_{\mu\nu}-\frac{1}{2}g_{\mu\nu}R = \kappa T_{\mu\nu}
\ , \label{Einstein}
\end{equation}
with 
$\kappa = 8\pi G$
and stress-energy tensor $T_{\mu\nu}$
\begin{eqnarray}
T_{\mu \nu} &=& \phantom{-} g_{\mu \nu} L_M 
-2 \frac{ \partial L}{\partial g^{\mu\nu}}
 \\
&=&-g_{\mu\nu} \left[ \frac{1}{2} g^{\alpha\beta} 
\left( \Phi_{, \, \alpha}^*\Phi_{, \, \beta}+
\Phi_{, \, \beta}^*\Phi_{, \, \alpha} \right)+U(\f)\right]
 + \left( \Phi_{, \, \mu}^*\Phi_{, \, \nu}
+\Phi_{, \, \nu}^*\Phi_{, \, \mu} \right
) \ .
\label{tmunu} \end{eqnarray}
Variation with respect to the scalar field
leads to the matter equation,
\begin{equation}
\left(\Box+\frac{\partial U}{\partial\left|\Phi\right|^2}\right)\Phi=0 \ ,
\label{field_phi}
\end{equation}
where $\Box$ represents the covariant d'Alembert operator. 
Equations (\ref{Einstein}) and (\ref{field_phi}) represent
the general set of non-linear Einstein--Klein--Gordon equations.

\subsection{Global Charges}\label{c1s3}

The mass $M$ and the angular momentum $J$
of stationary asymptotically flat space-times
can be obtained from their respective Komar expressions \cite{wald},
\begin{equation}
{M} = 
 \frac{1}{{4\pi G}} \int_{\Sigma}
 R_{\mu\nu}n^\mu\xi^\nu dV
\ , \label{komarM1}
\end{equation}
and
\begin{equation}
{\cal J} =  -
 \frac{1}{{8\pi G}} \int_{\Sigma}
 R_{\mu\nu}n^\mu\eta^\nu dV
\ . \label{komarJ1}
\end{equation}
Here $\Sigma$ denotes an asymptotically flat spacelike hypersurface,
$n^\mu$ is normal to $\Sigma$ with $n_\mu n^\mu = -1$,
$dV$ is the natural volume element on $\Sigma$,
$\xi$ denotes an asymptotically timelike Killing vector field
and $\eta$ an asymptotically spacelike Killing vector field
\cite{wald}.
Replacing the Ricci tensor via the Einstein equations by the
stress-energy tensor yields 
\begin{equation}
M
= \, 2 \int_{\Sigma} \left(  T_{\mu \nu} 
-\frac{1}{2} \, g_{\mu\nu} \, T_{\gamma}^{\ \gamma}
 \right) n^{\mu }\xi^{\nu} dV \ ,
 \label{komarM2}
\end{equation}
and
\begin{equation}
{\cal J} = -
 \int_{\Sigma} \left(  T_{\mu \nu}
-\frac{1}{2} \, g_{\mu\nu} \, T_{\gamma}^{\ \gamma}
 \right) n^{\mu }\eta^{\nu} dV \ .
 \label{komarJ2}
\end{equation}

A conserved charge $Q$ is associated
with the complex scalar field $\Phi$,
since the Lagrange density is invariant under the global phase transformation
\begin{equation}
\displaystyle
\Phi \rightarrow \Phi e^{i\alpha} \ ,
\end{equation}
leading to the conserved current
\begin{eqnarray}
j^{\mu} & = &  - i \left( \Phi^* \partial^{\mu} \Phi 
 - \Phi \partial^{\mu}\Phi ^* \right) \ , \ \ \
j^{\mu} _{\ ; \, \mu}  =  0 \ .
\end{eqnarray}

\section{Ansatz and Boundary Conditions}\label{c2}

\subsection{Ansatz}\label{c2s1}

To obtain stationary axially symmetric solutions,
we impose on the space-time the presence of
two commuting Killing vector fields,
$\xi$ and $\eta$, where
\begin{equation}
\xi=\partial_t \ , \ \ \ \eta=\partial_{\varphi}
\   \label{xieta} \end{equation}
in a system of adapted coordinates $\{t, r, \theta, \varphi\}$.
In these coordinates the metric is independent of $t$ and $\varphi$,
and can be expressed in isotropic coordinates
in the Lewis--Papapetrou form \cite{kkrot1}
\begin{eqnarray}
ds^2 &=&- f dt^2 
+ \frac{l}{f} \, \biggl[ h \left( dr^2 + r^2 \, d\t^2 \right) 
  + r^2 \, \sin^2 \t \,  \, \left( d \varphi
- \frac{\omega}{r} \, dt \right)^2 \biggr] \ . \label{ansatzg}
\end{eqnarray}
The four metric functions $f$, $l$, $h$ and $\omega$
are functions of the variables $r$ and $\theta$ only.

The symmetry axis of the spacetime, where $\eta=0$, 
corresponds to the $z$-axis.
The elementary flatness condition \cite{book}
\begin{equation}
\frac{X,_\mu X^{, \, \mu}}{4X} = 1 \ , \ \ \
X=\eta^\mu \eta_\mu \
\    \label{regcond} \end{equation}
then imposes on the symmetry axis the condition \cite{kk}
\begin{equation}
h|_{\theta=0}=h|_{\theta=\pi}=1 \ .
\end{equation}

For the scalar field $\Phi$ we adopt the stationary Ansatz \cite{schunck}
\begin{eqnarray}
\Phi (t,r,\t, \varphi)& = & \f (r, \t)
 e^{ i\omega_s t +i n \varphi} \ , \label{ansatzp}
\end{eqnarray}
where $\f (r, \t)$ is a real function,
and $\omega_s$ and $n$ are real constants.
Single-valuedness of the scalar field requires
\begin{equation}
\Phi(\varphi)=\Phi(2\pi + \varphi) \ , 
\end{equation}
thus the constant $n$ must be an integer,
i.e., $n \, = \, 0, \, \pm 1, \, \pm 2, \, \dots \ $.
We refer to $n$ as the rotational quantum number.
When $n \not=0$, the phase factor $\exp{(i n \varphi)}$ 
prevents spherical symmetry of the scalar field $\Phi$. 

Solutions with positive and negative parity satisfy, respectively,
\begin{eqnarray}
\f (r, \pi-\t) & = & {\phantom{-}} \f (r, \t) \\
\f (r, \pi-\t) & = &           -   \f (r, \t) 
\ , \label{parity}
\end{eqnarray}

To construct stationary axially symmetric boson star solutions
a system of five coupled partial differential equations
needs to be solved.
In contrast, for $Q$-balls 
the metric is the Minkowski metric, i.e.,$f=l=h=1$, $\omega=0$.
Here, at least in principle,
only a single partial differential equation for the scalar field
function needs to be solved.

\subsection{Mass, angular momentum and charge}\label{c2s2}

The mass $M$ and the angular momentum $J$ 
can be read off the asymptotic expansion of the metric functions $f$ 
and $\omega$, respectively,
\cite{kkrot1}
\begin{equation}
f = 1- \frac{2MG}{r} + O\left( \frac{1}{r^2} \right) \ , \ \ \
 \omega = \frac{2JG}{r^2} + O\left( \frac{1}{r^3} \right)\ ,
\label{MQasym1} \end{equation}
i.e.,
\begin{eqnarray}
M=\frac{1}{2G} \lim_{r \rightarrow \infty} r^2\partial_r \, f 
\ , \ \ \  J=\frac{1}{2G} \lim _{r \rightarrow \infty} r^2\omega \ .
\label{MJ2}
\end{eqnarray}
This is seen by considering the Komar expressions
Eqs.~(\ref{komarM1}) and (\ref{komarJ1}),
with unit vector $n^\mu = (1, 0, 0, \omega/r)/\sqrt{f}$,
and volume element 
$dV =1/ \sqrt{f} \, |g|^{1/2} \, dr \, d\t \, d\varphi$,
leading to \cite{kkrot2}
\begin{eqnarray}
 M &=&
 -\frac{1}{ 8 \pi G} \int_\Sigma  R_t^t \sqrt{-g} dr d\theta d\vphi 
\nonumber \\
 &=&  \lim_{r\to\infty}
 \frac{2\pi }{8\pi G} \int_0^{\pi}
 \left. \left[\frac{\sqrt{l}}{f}  r^2  \sin\theta
 \left( \frac{\partial f}{\partial r} -
 \frac{l}{f} \sin^2\theta \omega
 \left(\frac{\partial \omega}{\partial r} - \frac{\omega}{r}   \right)
 \right)
 \right]  \right|_{r} d\theta
 \ , \phantom{\frac{2\pi }{4\pi G}}
\end{eqnarray}
and similarly
\begin{equation}
J=\lim_{r\to\infty} 
\frac{2\pi }{16\pi G} \int_0^{\pi}
\left. \left[
\frac{l^{3/2}}{f^2} r^2 \sin^3\theta
\left( \omega - r \frac{\partial\omega}{\partial r} \right) 
\right]  \right|_{r} d\theta \ .
\label{localJ} \end{equation}
Insertion of the asymptotic expansions of the metric functions
then yields expressions (\ref{MJ2}). 

Alternatively, the mass $M$ and the angular momentum $J$
can be obtained by direct integration of the expressions
(\ref{komarM2}) and (\ref{komarJ2}), where
\begin{eqnarray}
M
&=& \phantom{2}\int_{\Sigma} \left( 2 T_{\e}^{\g} 
 - \d _{\e}^{\g} \, T_{\c}^{\c} \right) \, n_{\g} \, \xi^{\e} dV \ , 
\nonumber\\
&=&\int \left(2 \, T_t^t -T_{\e}^{\e} \right) \, |g|^{1/2} 
\, dr \, d\t \, d \varphi \ , \label{tolman}
\end{eqnarray}
corresponds to the Tolman mass, and
\begin{eqnarray}
J&=& -
\int T_{\, \varphi}^{\, t} \, |g|^{1/2} \, dr \, d\t \, d \varphi \ . 
\label{ang2}
\end{eqnarray}

The conserved scalar charge $Q$ is obtained from the time-component
of the current,
\begin{eqnarray}
Q &=- & \int j^t \left| g \right|^{1/2} dr d\t d\varphi 
\nonumber \\
 &=& 4 \pi \omega_s \int_0^{\infty}\int _0^{\pi} 
 |g| ^{1/2}   \frac{1}{f}  \left(  1 +
  \frac{n}{\omega_s}\frac{\omega}{r} \right) \f^2 \,
dr \, d\t 
\ . \label{Qc}
\end{eqnarray}

From Eq.~(\ref{ang2}) for the angular momentum $J$
and Eq.~(\ref{Qc}) for the scalar charge $Q$,
one obtains the important quantization relation for the angular momentum,
\begin{equation}
J=n \, Q \ , \label{JnQ}
\end{equation}
first derived by Schunck and Mielke \cite{schunck},
by taking into account that 
$T_{\, \varphi}^{\, t} = n j^t$, since
$\partial_\varphi \Phi =  i n \Phi$.
Thus a boson star with $n=0$ carries no angular momentum, $J =0$.

\subsection{Boundary conditions}\label{c2s3}

The choice of appropriate boundary conditions must guarantee 
that the boson star solutions
are globally regular and asymptotically flat,
and that they possesses finite energy and finite energy density.

For rotating axially symmetric boson stars
appropriate boundary conditions must be specified
for the metric functions $f(r,\theta)$, $l(r,\theta)$, $h(r,\theta)$
and $\omega(r,\theta)$, 
and the scalar field function $\f(r,\theta)$
at infinity, at the origin, on the positive $z$-axis ($\theta=0$),
and, exploiting the reflection properties w.r.t.~$\theta \rightarrow
\pi - \theta$, in the $xy$-plane ($\theta=\pi/2$).

For $r \rightarrow \infty$ 
the metric approaches the Minkowski metric $\eta_{\a\b}$
and the scalar field assumes its vacuum value $\Phi=0$. 
Accordingly, we impose at infinity the boundary conditions 
\begin{equation}
f|_{r \rightarrow \infty} =1 \ , \ \ \
l|_{r \rightarrow \infty} =1 \ , \ \ \
h|_{r \rightarrow \infty} =1 \ , \ \ \
\omega|_{r \rightarrow \infty} =0 \ , \ \ \
\f| _{r \rightarrow \infty}=0 \ .
\label{bc4} \end{equation}

At the origin we require
\begin{equation}
\partial_r f|_{r=0}=0 \ , \ \ \
\partial_r l|_{r=0}=0 \ , \ \ \
h|_{r=0}=1 \ , \ \ \
\omega|_{r=0}=0 \ , \ \ \
\f| _{r =0}=0 \ .
\label{bc3} \end{equation}
Note, that for spherically symmetric boson stars the scalar field
has a finite value $\f_0$ at the origin, though.

For $\t=0$ and $\t=\pi/2$, respectively, 
we require the boundary conditions
\begin{equation}
\partial_{\t} f|_{\t=0}=0 \ , \ \ \
\partial_{\t} l|_{\t=0}=0 \ , \ \ \
h|_{\t=0}=1 \ , \ \ \
\partial_{\t} \omega |_{\t=0}=0 \ , \ \ \
\f |_{\t=0}=0 \ , 
\label{bc5} \end{equation}
and
\begin{equation}
\partial_{\t} f|_{\t=\pi/2}=0 \ , \ \ \
\partial_{\t} l|_{\t=\pi/2}=0 \ , \ \ \
\partial_{\t} h|_{\t=\pi/2}=0 \ , \ \ \
\partial_{\t} \omega |_{\t=\pi/2}=0 \ , \ \ \
\partial_{\t} \f |_{\t=\pi/2}=0 \ ,
\label{bc6} \end{equation}
for even parity solutions,
while for odd parity solutions
$\f |_{\t=\pi/2}=0$.

In non-rotating solutions, the metric function $\omega$
is identically zero.
In spherically symmetric solutions
the functions depend on the radial coordinate only,
with metric function $h=1$.
For the flat space solutions the metric corresponds
to the Minkowski metric everywhere.

\subsection{Numerical method}

Rotating solutions are obtained when $n \ne 0$.
The resulting set of coupled non-linear partial differential equations
is solved numerically \cite{FIDISOL} 
subject to the above boundary conditions, Eqs.~(\ref{bc4})-(\ref{bc6}).
Because of the power law fall-off of the metric functions,
we compactify space by introducing the compactified radial coordinate
\begin{equation}
\bar r = \frac{r}{1+r} \ .
\label{rcomp} \end{equation}
The numerical calculations are based on the Newton-Raphson method.
The equations are discretized on a non-equidistant grid in
$\bar r$ and  $\theta$.
Typical grids used have sizes $90 \times 70$,
covering the integration region
$0\leq \bar r\leq 1$ and $0\leq\theta\leq\pi/2$.

\section{$Q$-balls}\label{c3}

\subsection{Spherically symmetric Q-balls}

We first briefly review the main features 
of spherically symmetric $Q$-balls 
(see e.g.~\cite{lee-s,coleman,volkov}, or \cite{lee-rev} for a review).
%
The equation of motion for
spherically symmetric $Q$-balls is given by
\begin{eqnarray}
\label{radeom}
0 & = & \phi'' + \frac{2} {r} \, \phi' -
\frac{1}{2} \frac{d U(\phi)} {d\phi} +
\omega_s^2 \, \phi \ .
\end{eqnarray}
It may be interpreted to
effectively describe a particle moving with friction in the 
effective potential
\begin{eqnarray}
V(\phi) = \frac{1}{2} \, \omega_s^2 \, \phi^2
 - \frac{1}{2} U(\phi) \ ,
\label{Vpot}
\end{eqnarray}
exhibited in Fig.~\ref{potentiale}.

$Q$-balls exist in the frequency range 
\begin{equation}
\omega_{\rm min}^2 < \omega_s^2 < \omega_{\rm max}^2 \ ,
\label{olimits}
\end{equation}
where the maximal frequency $\omega_{\rm max}$
and the minimal frequency $\omega_{\rm min}$ are determined
by the properties of the potential $U$ \cite{volkov},
\begin{eqnarray}
\label{cond1}
\omega_s^2 & < & \omega^{2}_{\rm max} \; \equiv \; 
\frac{1}{2} U''(0) = \lambda \, b \ ,
\end{eqnarray}
\begin{eqnarray}
\label{cond2}
\omega_s^2 & > & \omega^{2}_{\rm min} \; \equiv \; 
\min_{\f} \left[{U(\phi)}/{\phi^2} \right] \; = \; 
\lambda \left(b- \frac{a^2}{4} \right) \ .
\end{eqnarray}

The mass $M$ of the $Q$-balls
can be obtained from 
\begin{eqnarray}
\label{radenergy}
M
=
4 \pi \int_0^\infty  \left[\omega_s^2 \, \phi^2 +
\phi^{\, \prime \, 2} + U(\phi)\right] r^2 \, d r \ ,
\end{eqnarray}
where the prime denotes differentiation with respect to $r$, and
and their charge $Q$ from 
\begin{eqnarray}
\label{charge}
Q(\omega_s) & = & 8 \pi \, \omega_s \int_0^\infty \phi^2 \, r^2 \, d r \ .
\end{eqnarray}

Fixing the value of $\omega_s$ in the allowed range, 
one then obtains a sequence of $Q$-ball solutions, 
consisting of the fundamental $Q$-ball and its
radial excitations \cite{volkov}.
The boson function $\f$ of the fundamental $Q$-ball has no nodes,
while it has $k$ nodes for the $k$-th radial excitation.

\begin{figure}[h!]
\parbox{\textwidth}
{\centerline{
\mbox{
\epsfysize=10.0cm
\includegraphics[width=70mm,angle=0,keepaspectratio]{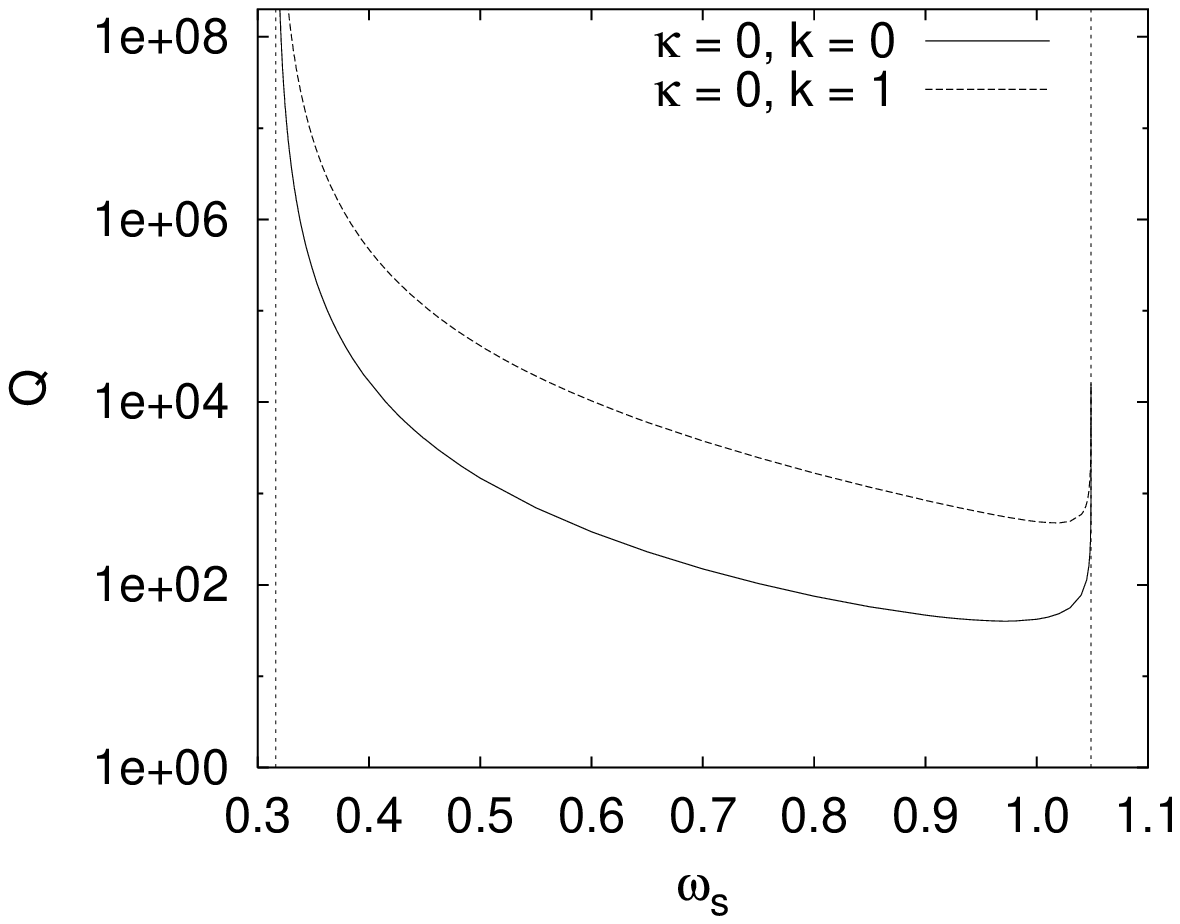}
\includegraphics[width=75mm,angle=0,keepaspectratio]{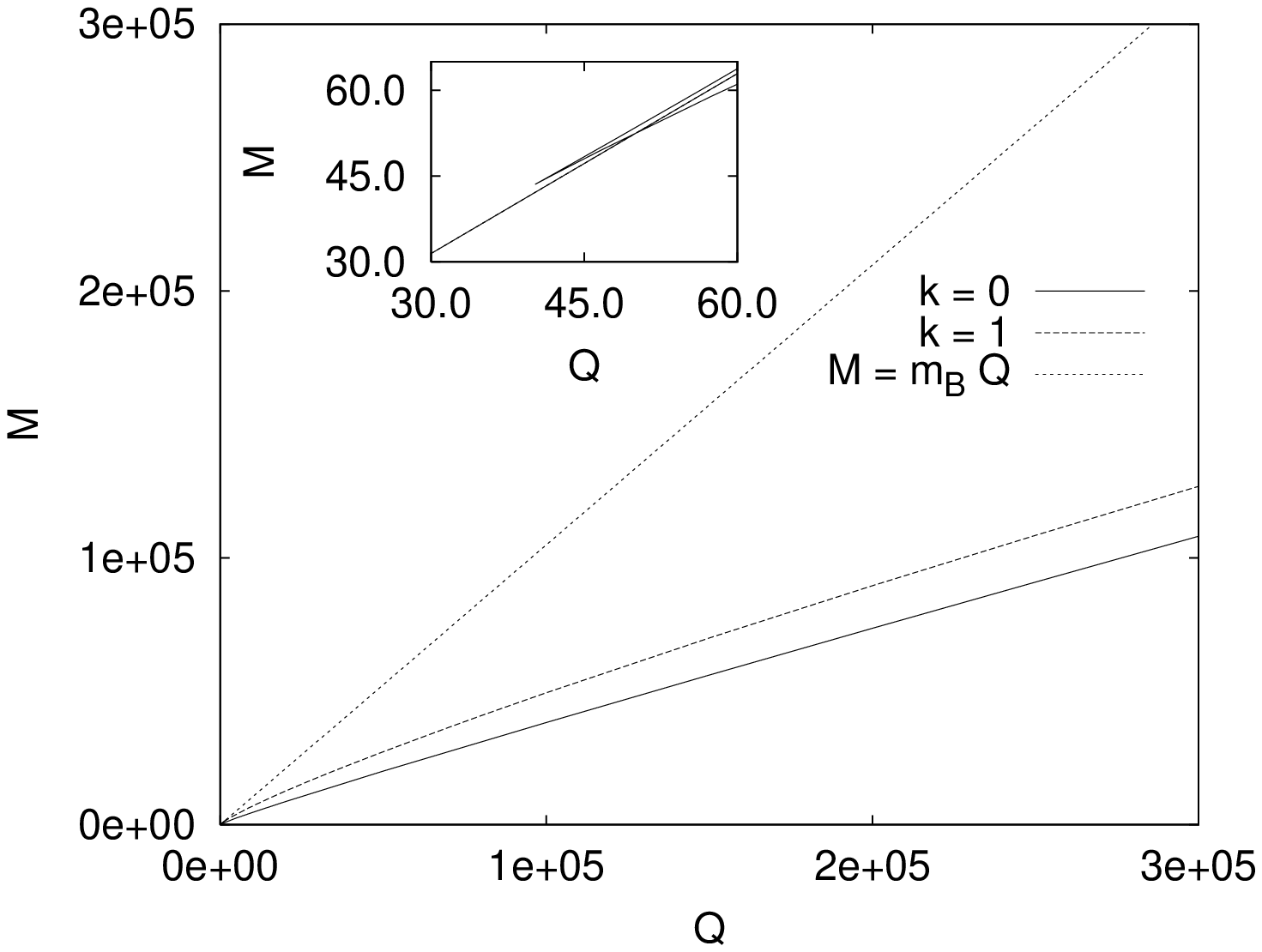}
}}}
\caption{
Left: The charge $Q$ versus the frequency $\omega_s$
for spherically symmetric fundamental $Q$-balls ($k=0$) 
and their first radial excitations ($k=1$).
Also shown are the limiting values of the frequency, $\omega_{\rm min}$
and $\omega_{\rm max}$.
Right: The mass $M$ versus the charge $Q$
for spherically symmetric fundamental $Q$-balls ($k=0$) 
and their first radial excitations ($k=1$).
Also shown is the mass for $Q$ free bosons, $M=m_{\rm B}Q$.
The mass on the upper branches is very close this mass.
The $k=0$ branches close to the critical
charge $Q_{\rm cr}$ are seen in the inset.
}
\label{Qvsomega0}
\end{figure}

We exhibit the fundamental $Q$-ball solution ($k=0$) and the first
radial excitation ($k=1$) in Fig.~\ref{Qvsomega0}.
Here the charge $Q$ is shown versus the frequency $\omega_s$.
As seen in the figure, at a critical value
$\omega_{\rm cr}$, which depends on $k$,
the charge assumes its respective minimal value $Q_{\rm cr}$.
Thus two branches of solutions exist for each $k$, 
the lower branch $\omega_s < \omega_{\rm cr}$
and the upper branch $\omega_s > \omega_{\rm cr}$,
which merge and end at $\omega_{\rm cr}$.
When $\omega_s \rightarrow \omega_{\rm min}$
as well as when $\omega_s \rightarrow \omega_{\rm max}$ 
the charge diverges \cite{lee-s}.

In Fig.~\ref{Qvsomega0} also the mass $M$ of these two sets of solutions
is shown, but it is now exhibited versus the charge $Q$.
The lower mass branches correspond to the lower values of the frequency,
$\omega_s < \omega_{\rm cr}$, while the upper mass branches
correspond to higher values of the frequency, $\omega_s > \omega_{\rm cr}$.
As long as the mass is smaller than the mass of $Q$ free bosons,
$M<m_{\rm B}Q$, the fundamental $Q$-ball solutions are stable,
both classically and quantum mechanically \cite{lee-s}.
Classically the solutions remain stable all along the lower branch.
The classical stability changes only at the critical point, $Q_{\rm cr}$,
where the solutions acquire an unstable mode \cite{lee-s}.
In contrast, from a quantum perspective the stability changes
when the lower mass branch intersects the line corresponding to
the mass of $Q$ free bosons.
On the unstable upper branch
the mass approaches the mass $M=m_{\rm B}Q$ of $Q$ free bosons from above,
as $\omega_s \rightarrow \omega_{\rm max}$ \cite{lee-s}.

We remark, that also angularly excited 
non-rotating $Q$-balls have been obtained recently \cite{Brihaye:2007tn}.
These new solutions are axially symmetric,
and their angular dependence is given by spherical harmonics. 
They have higher energy and charge than the spherically 
symmetric $Q$-balls, but lower energy and charge than the rotating $Q$-balls.
In addition, interacting Q-balls have been studied \cite{Brihaye:2007tn}.
 
\boldmath
\subsection{Rotating $Q$-balls}
\unboldmath

The existence of rotating $Q$-balls was shown by
Volkov and W\"ohnert \cite{volkov}.
Based on the Ansatz (\ref{ansatzp}) for the scalar field $\Phi$ \cite{schunck},
rotating $Q$-balls are solutions of the field equation
\begin{eqnarray}
\label{eom3d}
\left( \frac{\partial^2} {\partial r^2} + \frac{2} {r} \frac{\partial}
{\partial r} + \frac{1} {r^2} \frac{\partial^2} {\partial \theta^2}
+ \frac{\cos\theta} {r^2 \sin\theta} \frac{\partial} {\partial
\theta} - \frac{n^2} {r^2 \sin^2\theta} + \omega_s^2 \right) \phi &
= & \frac{1}{2} \frac{d U(\phi)}{d \phi} \ , \phantom{abcd}
\end{eqnarray}
with mass \cite{volkov}
\begin{equation}
\label{erot}
M = 2 \pi \int_0^\infty \int_0^\pi 
\left(\omega_s^2 \phi^2 + (\partial_r \phi)^2 +
\frac{1} {r^2} (\partial_\theta \phi)^2 + \frac{n^2\phi^2} {r^2
\sin^2 \theta} + U(\phi) \right) 
r^2 \, d r \, \sin\theta\, d \theta
\ , 
\end{equation}
and charge
\begin{equation}
Q= 4 \pi \omega_s \int_0^{\infty}\int _0^{\pi} 
 \f^2 \, r^2 dr \, \sin \theta \, d\t \ 
\ . \label{3} \end{equation}
Their angular momentum satisfies the quantization relation
$J=nQ$.

In their pioneering study \cite{volkov}
Volkov and W\"ohnert 
showed, that for a given value of $n$ there are two types of solutions,
positive parity $Q$-balls $n^+$ and negative parity $Q$-balls $n^-$.
In the following we restrict our considerations
to fundamental rotating $Q$-balls.

\subsubsection{Positive parity $Q$-balls}

\begin{figure}[t!]
\parbox{\textwidth}
{\centerline{
\mbox{
\epsfysize=10.0cm
\includegraphics[width=70mm,angle=0,keepaspectratio]{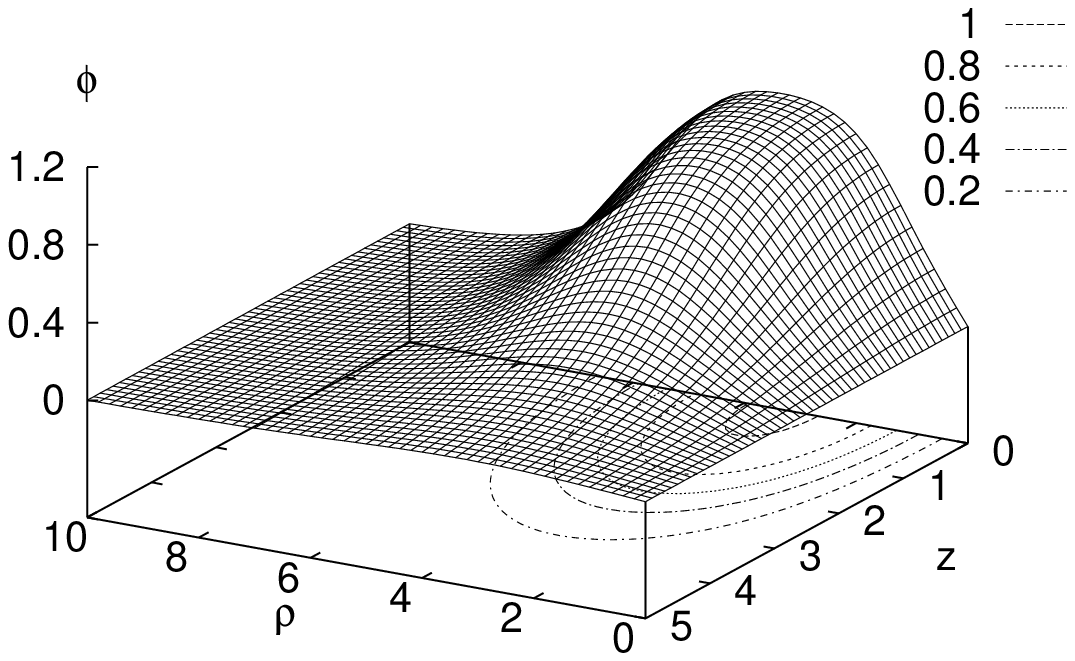}
\includegraphics[width=70mm,angle=0,keepaspectratio]{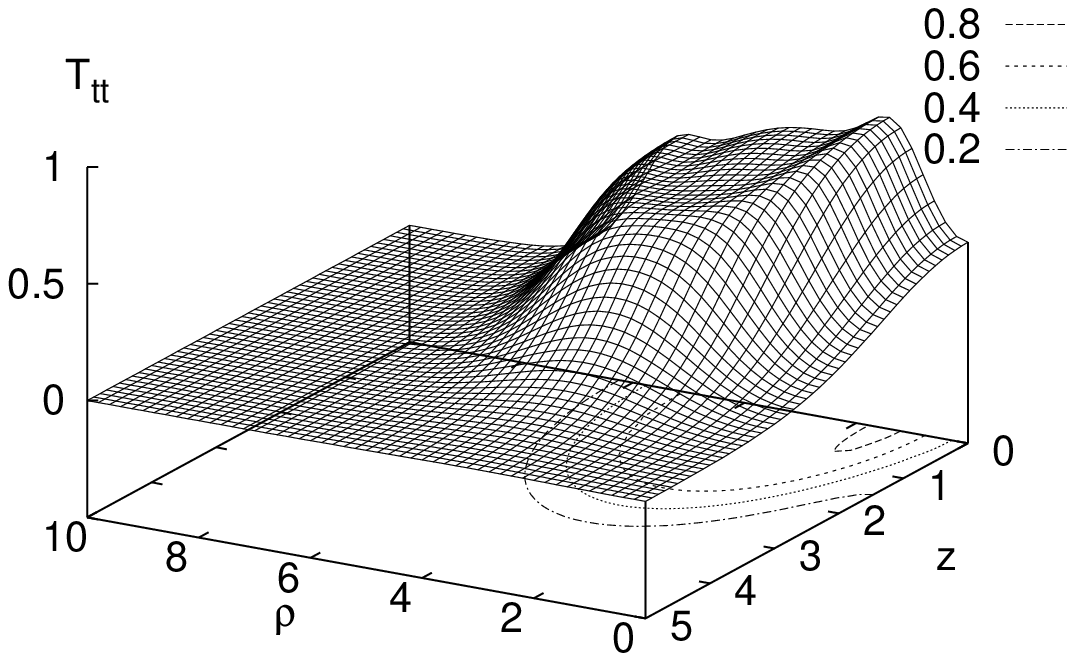}
}}}
{\centerline{
\mbox{
\epsfysize=10.0cm
\includegraphics[width=70mm,angle=0,keepaspectratio]{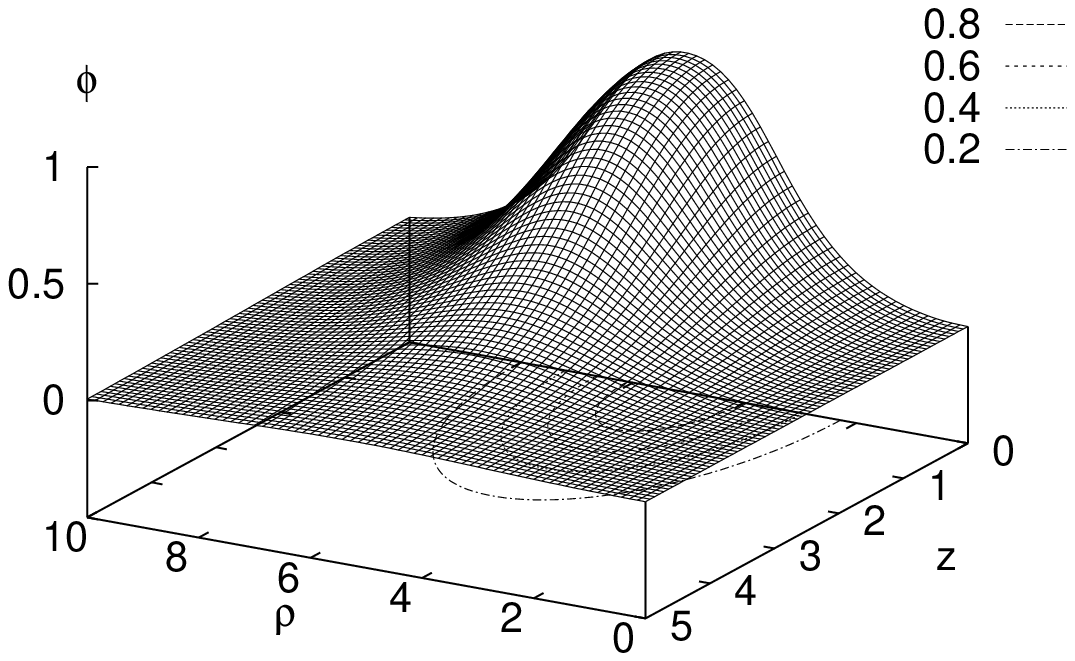}
\includegraphics[width=70mm,angle=0,keepaspectratio]{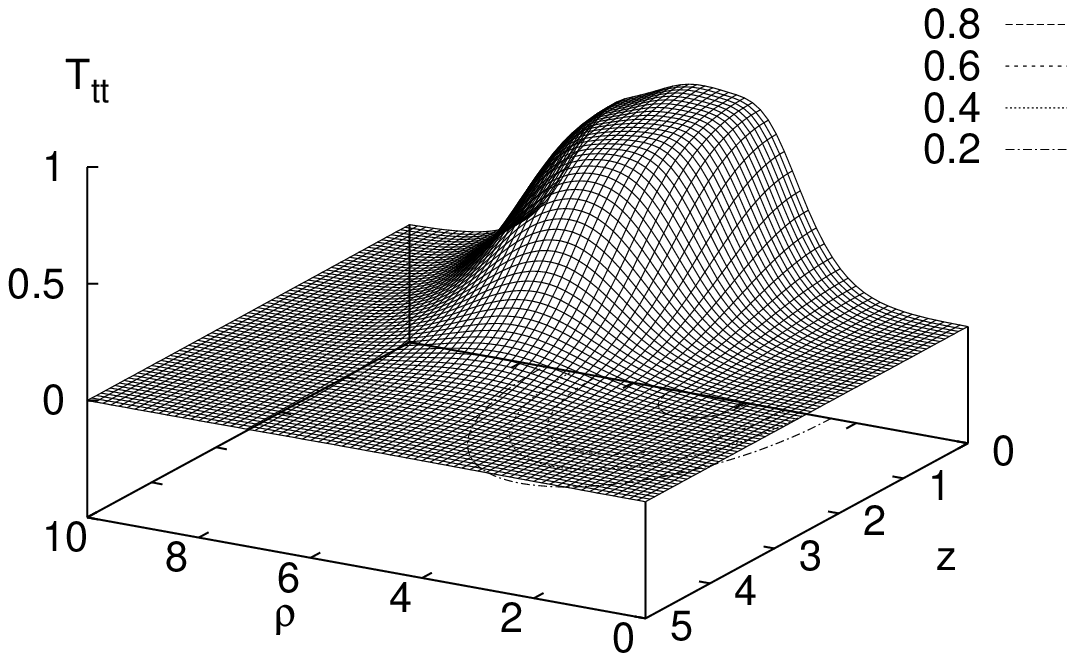}
}}}
{\centerline{
\mbox{
\epsfysize=10.0cm
\includegraphics[width=70mm,angle=0,keepaspectratio]{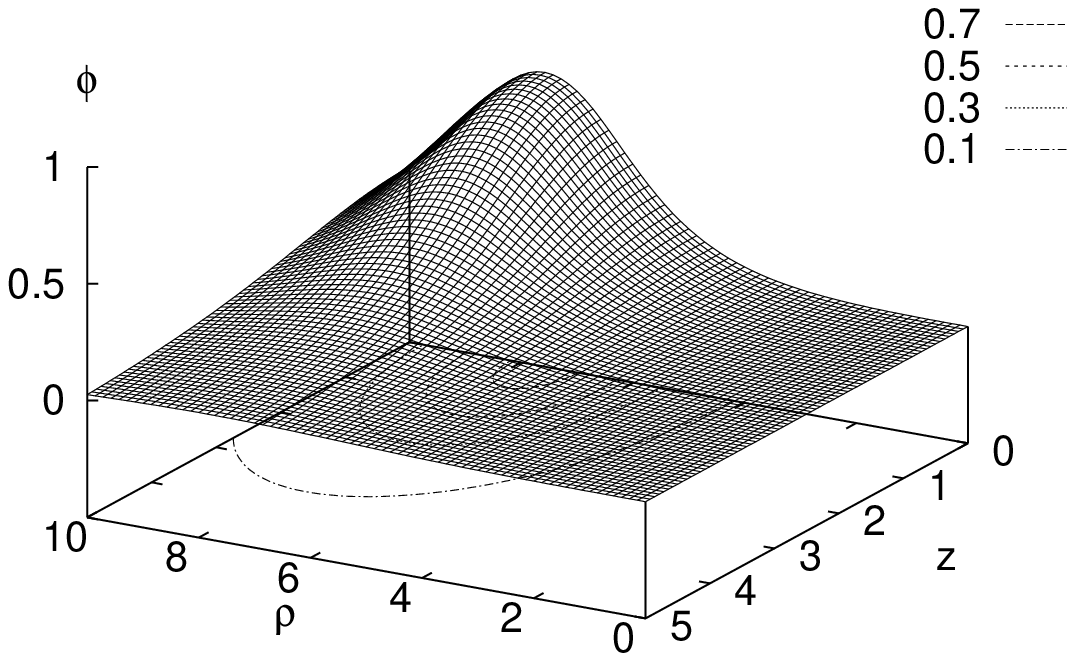}
\includegraphics[width=70mm,angle=0,keepaspectratio]{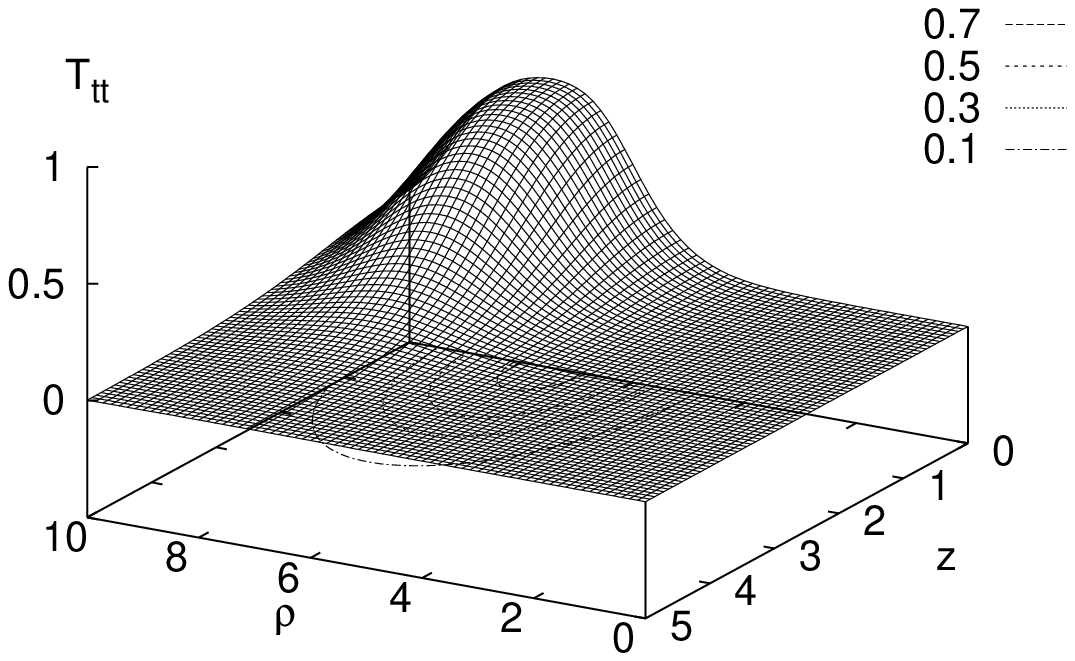}
}}}
\caption{
The scalar field $\f$ (left column)
and energy density $T_{tt}$ (right column)
for rotating positive parity $Q$-balls
with charge $Q=410$ and 
$n^P=1^+$ (upper row), $2^+$ (middle row) and $3^+$ (lower row)
versus the coordinates $\rho=r \sin \theta$
and $z= r \cos\theta$ for $z\ge 0$.
}
\label{QT410_n2}
\end{figure} 

We first recall the properties of
rotating $Q$-balls with positive parity
and, at the same time, extend the previous studies \cite{volkov,list}
to higher rotational quantum number $n$.
In Fig.~\ref{QT410_n2}
we illustrate the scalar field $\f$ and the energy density
$T_{tt}$ of rotating $Q$-balls with charge $Q=410$,
$n^P=1^+$, $2^+$ and $3^+$.
For rotating $Q$-balls the scalar field $\f$ vanishes on the
$z$-axis.
The energy density $T_{tt}$ of even parity rotating $Q$-balls 
typically exhibits a torus-like structure.
This torus-like shape becomes apparent by
considering surfaces of constant energy density
of e.g.~half the respective maximal value of the energy density.
As seen in the figure, the maximum of the energy density 
increases and shifts towards
larger values of the coordinate $\rho=r \sin \theta$,
when $n$ and thus the angular momentum increases.
In terms of the energy density tori this manifests as an
increase of their radii in the equatorial plane with increasing $n$.
(Note, that in \cite{list} a weighted energy density was shown.)

\begin{figure}[h!]
\parbox{\textwidth}
{\centerline{
\mbox{
\epsfysize=10.0cm
\includegraphics[width=70mm,angle=0,keepaspectratio]{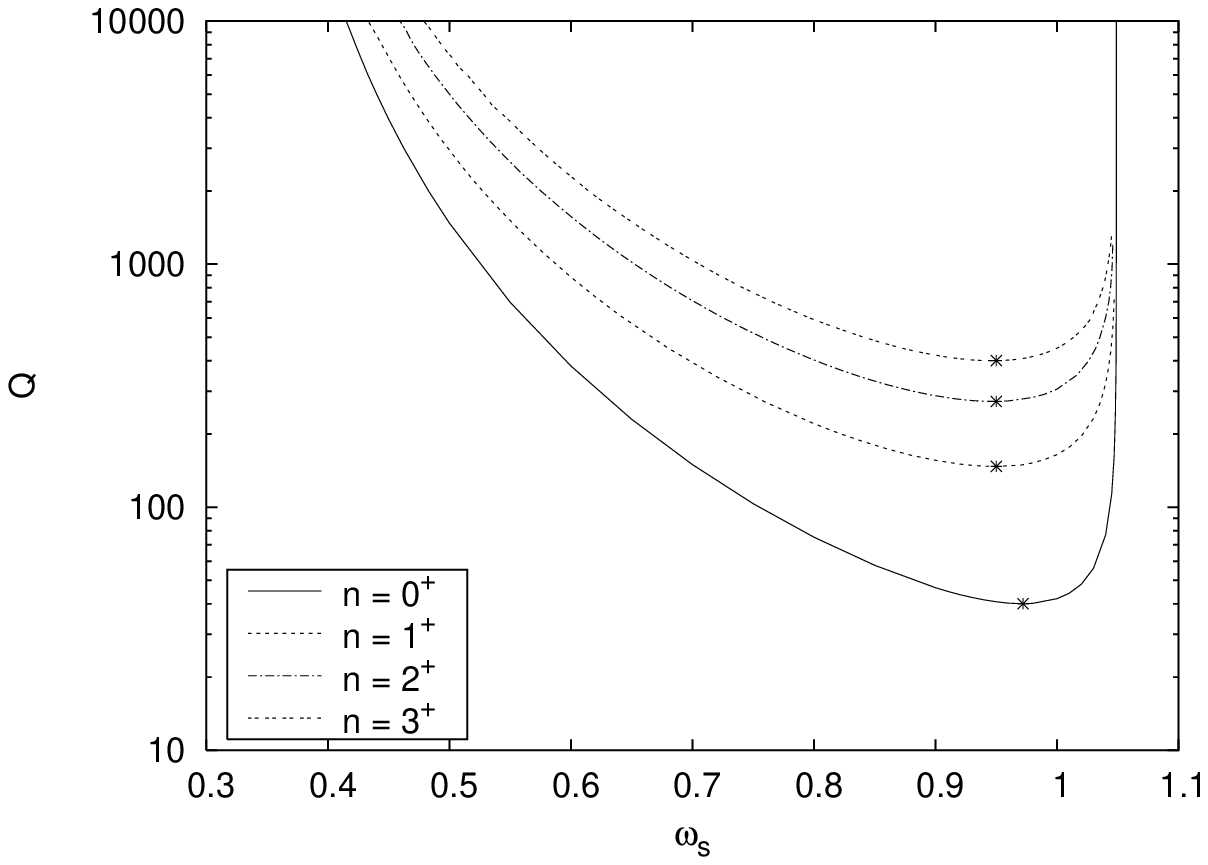}
\includegraphics[width=70mm,angle=0,keepaspectratio]{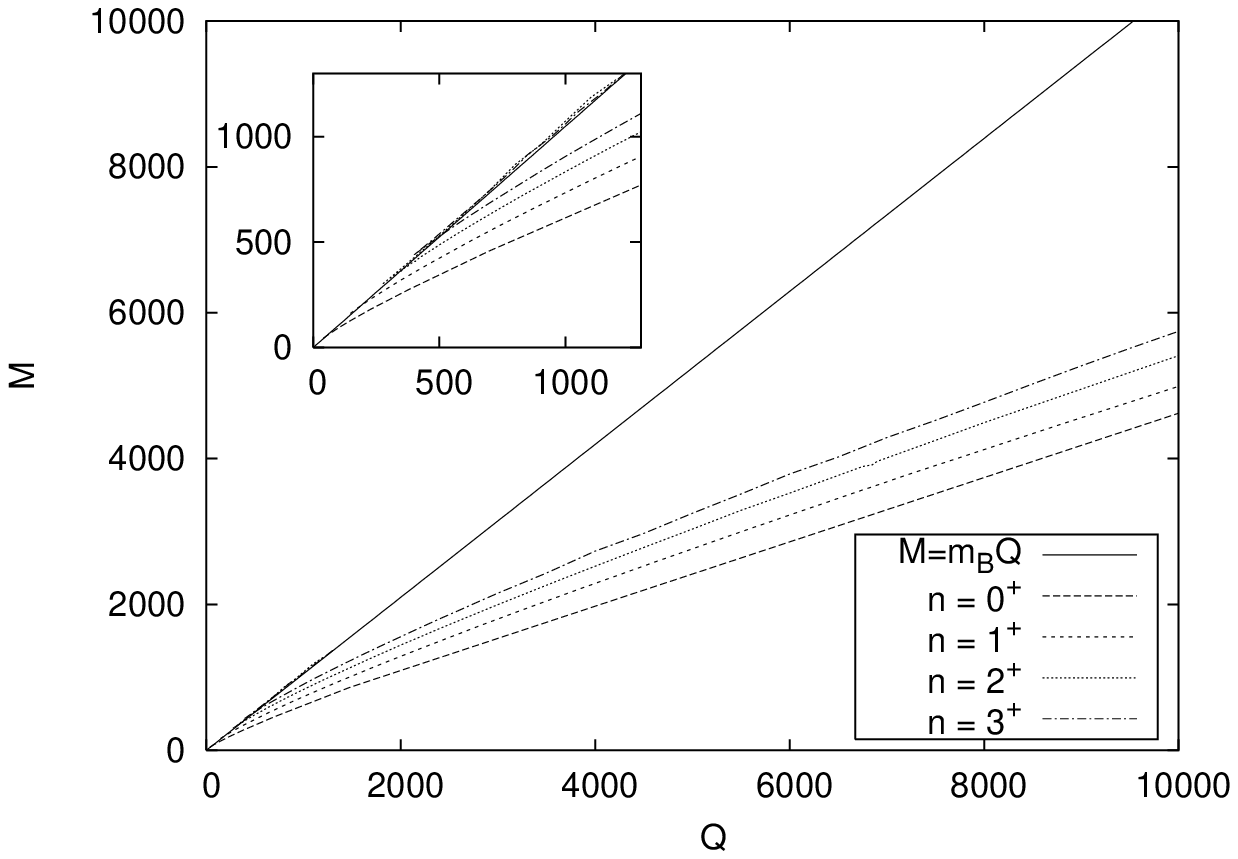}
}}}
\caption{
Left: The charge $Q$ versus the frequency $\omega_s$
for $Q$-balls with 
$n^P=0^+$, $1^+$, $2^+$, $3^+$.
The asterisks mark the critical values of the charge.
Right: The mass $M$ versus the charge $Q$
for $Q$-balls with 
$n^P=0^+$, $1^+$, $2^+$, $3^+$.
The upper branches of the mass $M$ are hardly discernible (on this scale)
from the mass of $Q$ free bosons, $M=m_{\rm B}Q$.
}
\label{rot-flat}
\end{figure}

We next address the frequency dependence for these 
rotating $Q$-ball solutions with positive parity.
In Fig.~\ref{rot-flat} we exhibit the charge $Q$ 
versus the frequency $\omega_s$ 
for $Q$-balls with $n=0-3$.
We observe the same upper limiting value $\omega_{\rm max}$, Eq.~(\ref{cond1}),
for the frequency $\omega_s$, as for non-rotating $Q$-balls,
ensuring asymptotically an exponential fall-off of the scalar field $\f$.
For a given frequency $\omega_s$ the charge of a 
$Q$-ball increases with $n$.
We thus conclude, that the frequency of rotating $Q$-balls
is limited by the minimal frequency
$\omega_{\rm min}$, Eq.~(\ref{cond2}), as well,
independent of $n$.

The mass of these rotating $Q$-balls with positive parity
shows the same critical behaviour and thus cusp structure
as the mass of the non-rotating $Q$-balls. 
This is also illustrated in Fig.~\ref{rot-flat}.
Thus we conclude, that these rotating solutions are classically stable
along their lower mass branches.
Overall we note, that the sets of rotating $Q$-balls with positive parity
exhibit the same general pattern as the sets of non-rotating $Q$-balls.

\subsubsection{Negative parity $Q$-balls}

\begin{figure}[h!]
\parbox{\textwidth}
{\centerline{
\mbox{
\epsfysize=10.0cm
\includegraphics[width=70mm,angle=0,keepaspectratio]{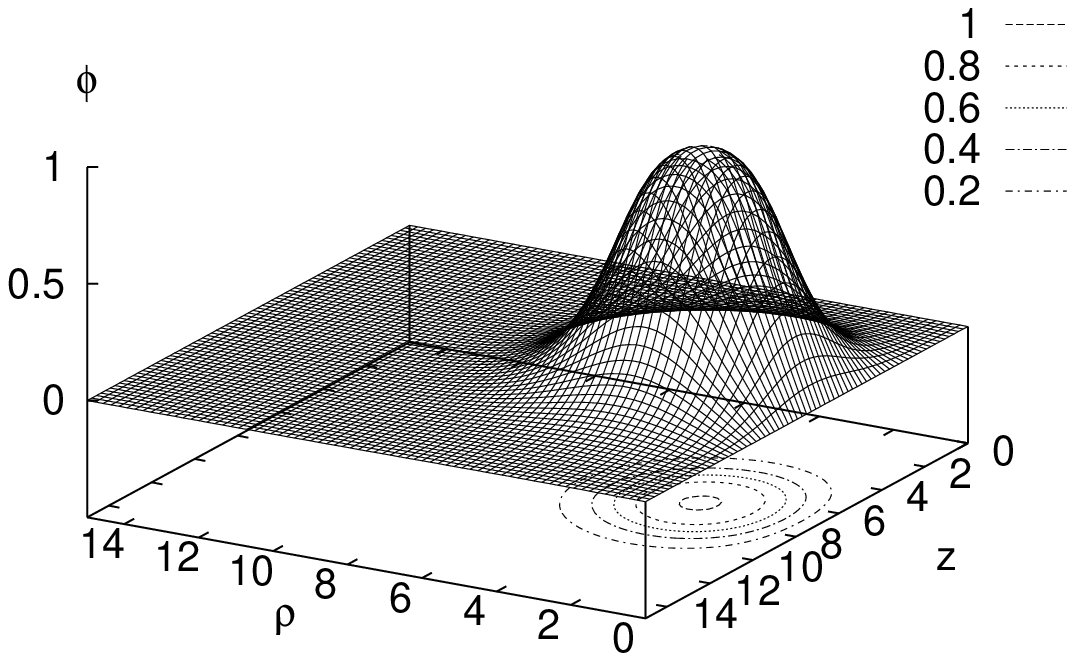}
\includegraphics[width=70mm,angle=0,keepaspectratio]{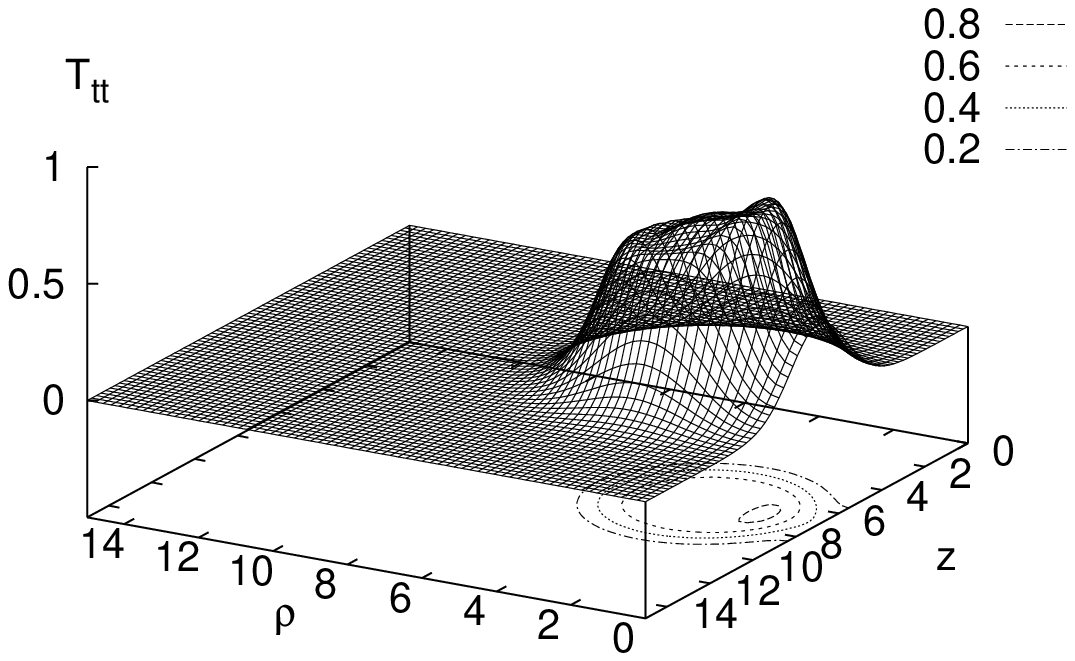}
}}}
{\centerline{
\mbox{
\epsfysize=10.0cm
\includegraphics[width=70mm,angle=0,keepaspectratio]{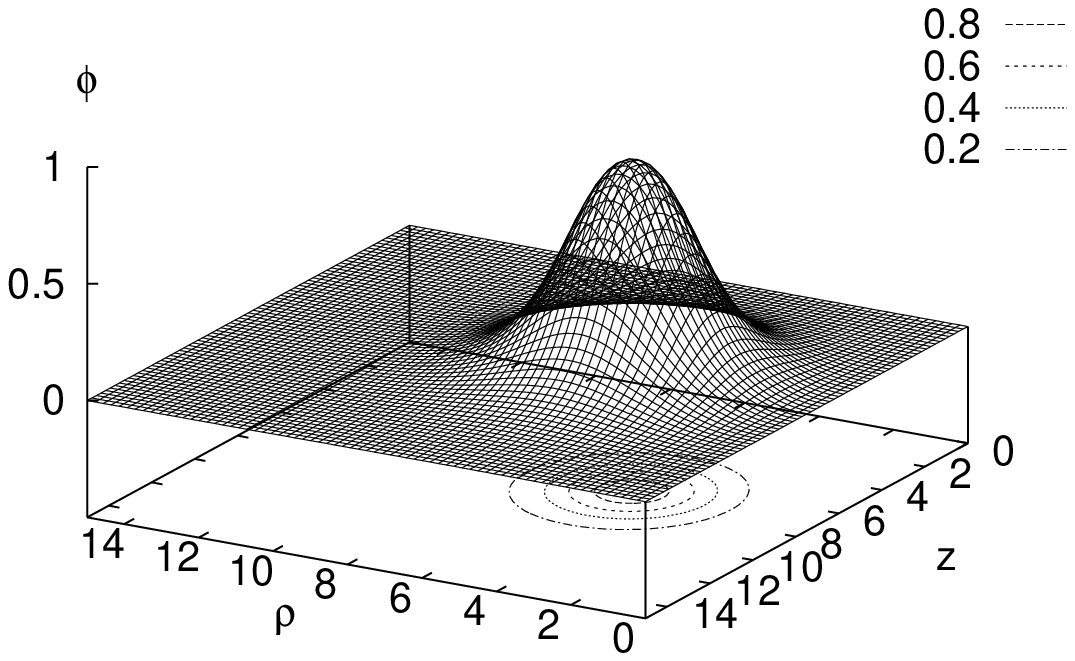}
\includegraphics[width=70mm,angle=0,keepaspectratio]{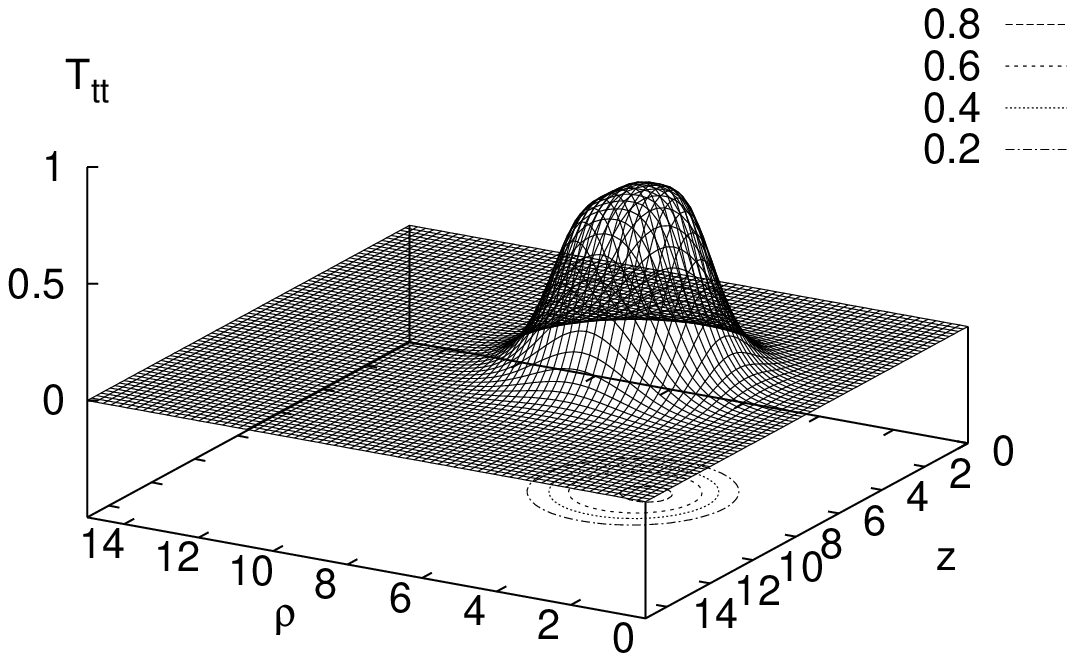}
}}}
\caption{
The scalar field $\f$ (left column)
and energy density $T_{tt}$ (right column)
for rotating negative parity $Q$-balls
with charge $Q=720$ and 
$n^P=1^-$ (upper row) and $2^-$ (lower row)
versus the coordinates $\rho=r \sin \theta$
and $z= r \cos\theta$ for $z\ge 0$.
}
\label{QT410_n3}
\end{figure}

Turning to rotating $Q$-balls with negative parity,
we first illustrate in Fig.~\ref{QT410_n3}
the scalar field $\f$ and the energy density
$T_{tt}$ of $Q$-balls with 
fixed charge $Q=720$
and 
$n^P=1^-$ and $2^-$.
For these $Q$-balls the scalar field $\f$ vanishes not only 
on the $z$-axis but everywhere in the $xy$-plane, as well.
The energy density $T_{tt}$ now typically
exhibits a double torus-like structure,
where the two tori are located symmetrically
w.r.t.~the equatorial plane
\cite{volkov}.
When $n$ and thus the angular momentum increases,
the maxima of the energy density again shift towards
larger values of the coordinate $\rho=r \sin \theta$.
For the double tori of the energy density this shift 
with increasing $n$ manifests as an
increase of the radii of the tori in the equatorial plane.

\begin{figure}[h!]
\parbox{\textwidth}
{\centerline{
\mbox{
\epsfysize=10.0cm
\includegraphics[width=70mm,angle=0,keepaspectratio]{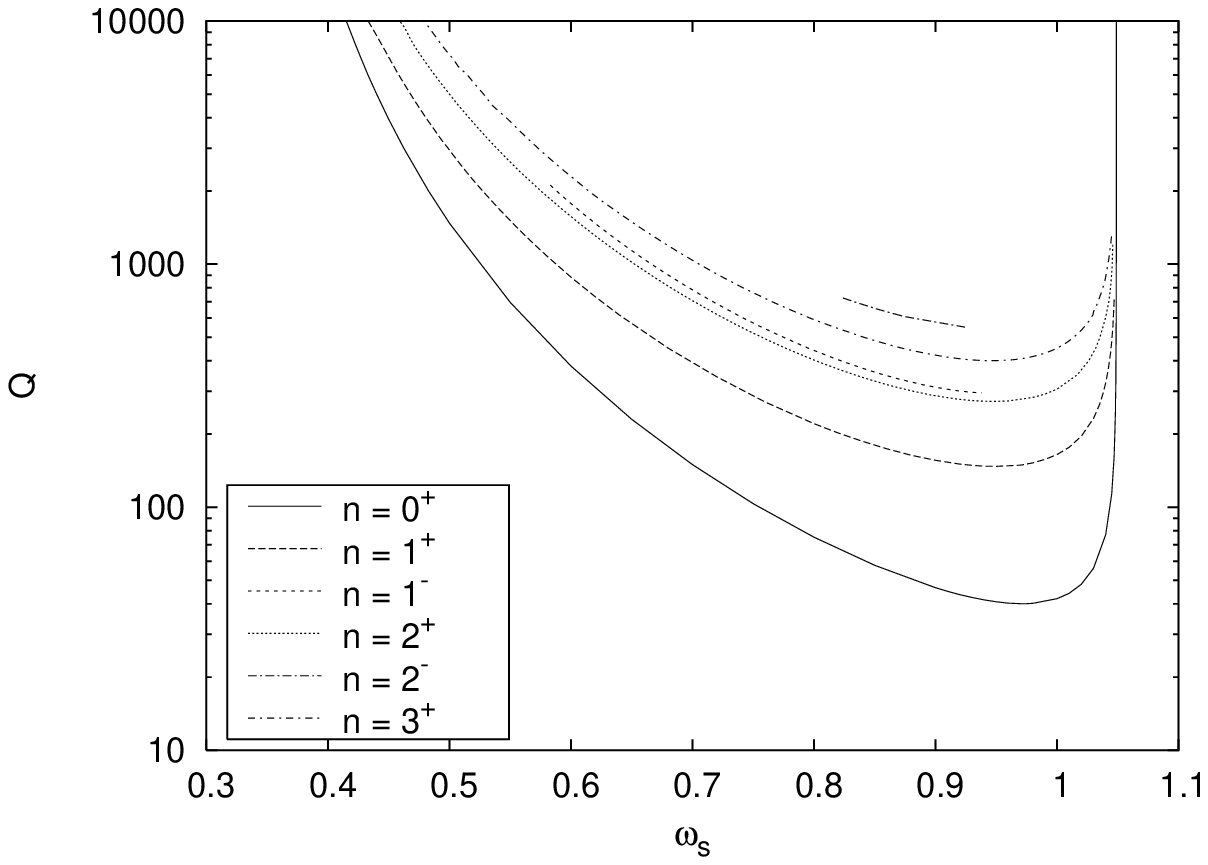}
\includegraphics[width=70mm,angle=0,keepaspectratio]{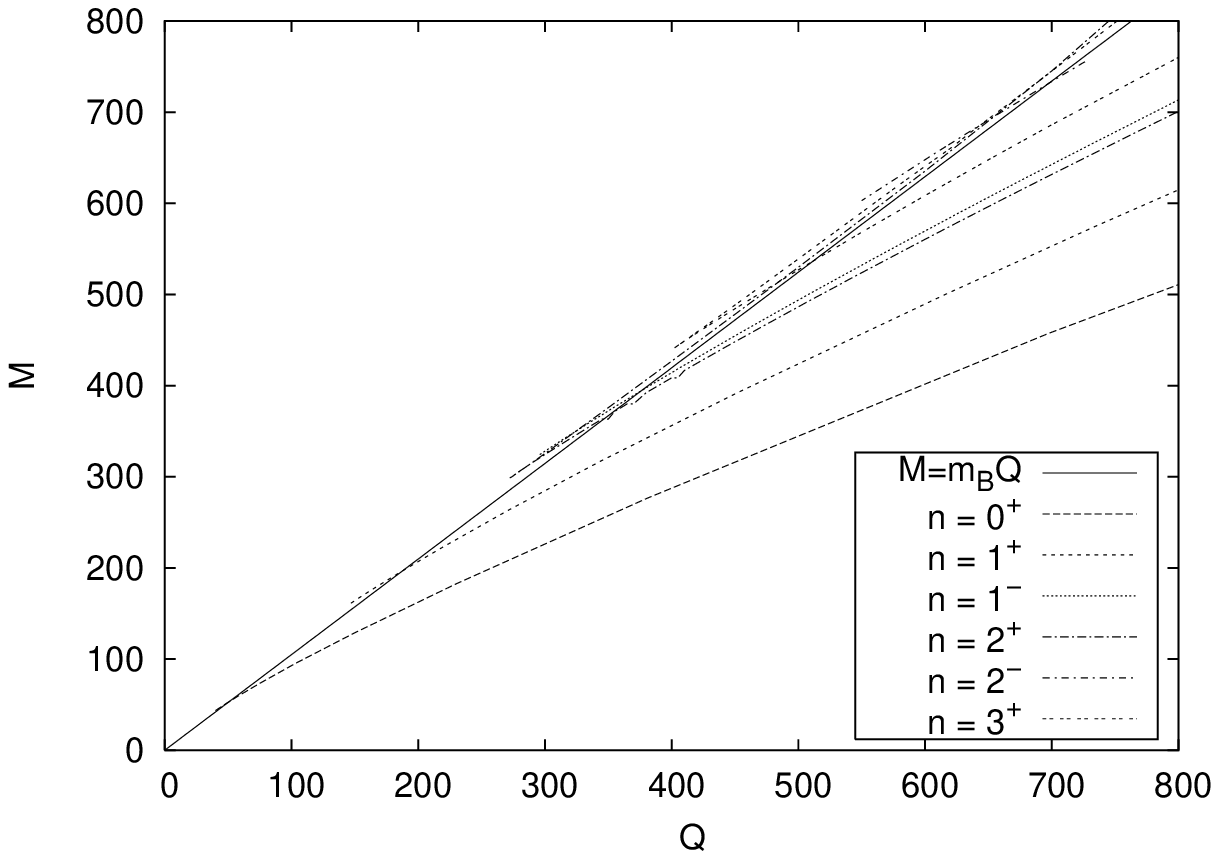}
}}}
\caption{
Left: The charge $Q$ versus the frequency $\omega_s$
for negative parity $Q$-balls with $n^P=1^-$ and $2^-$.
For comparison the positive parity $Q$-balls 
with $n^P=0^+$, $1^+$, $2^+$, $3^+$ are also shown.
Right: The mass $M$ versus the charge $Q$
for the same set of solutions together with
the mass of $Q$ free bosons, $M=m_{\rm B}Q$.
}
\label{rot-flat2}
\end{figure}

In Fig.~\ref{rot-flat2} we exhibit the charge $Q$ 
versus the frequency $\omega_s$ 
for negative parity $Q$-balls with 
$n^P=1^-$ and $2^-$, and compare to
the positive parity $Q$-balls.
While we conclude, that
for the negative parity $Q$-balls
the limiting values $\omega_{\rm min}$ and $\omega_{\rm max}$
are retained,
numerical difficulties have unfortunately prevented a more complete study.

Comparing the sets of negative and positive parity $Q$-balls,
we conclude,
that they exhibit the same general pattern.
However, 
concerning the magnitudes of their charges and masses,
the positive parity $Q$-balls and the negative parity $Q$-balls
do not quite alternate. 
Instead, they are ordered according to
$n^P=0^+,\, 1^+,\, 2^+,\, 1^-,\, 3^+,\, 2^-$, as seen in
Fig.~\ref{rot-flat2}.
Thus for $Q$-balls of a given charge $Q$,
the double torus-like structure of the negative parity solutions
leads to a considerably higher mass
than the single torus-like structure of the positive parity solutions.

\section{Boson stars}

\subsection{Spherically symmetric boson stars}

\begin{figure}[h!]
\parbox{\textwidth}
{\centerline{
\mbox{
\epsfysize=10.0cm
\includegraphics[width=80mm,angle=0,keepaspectratio]{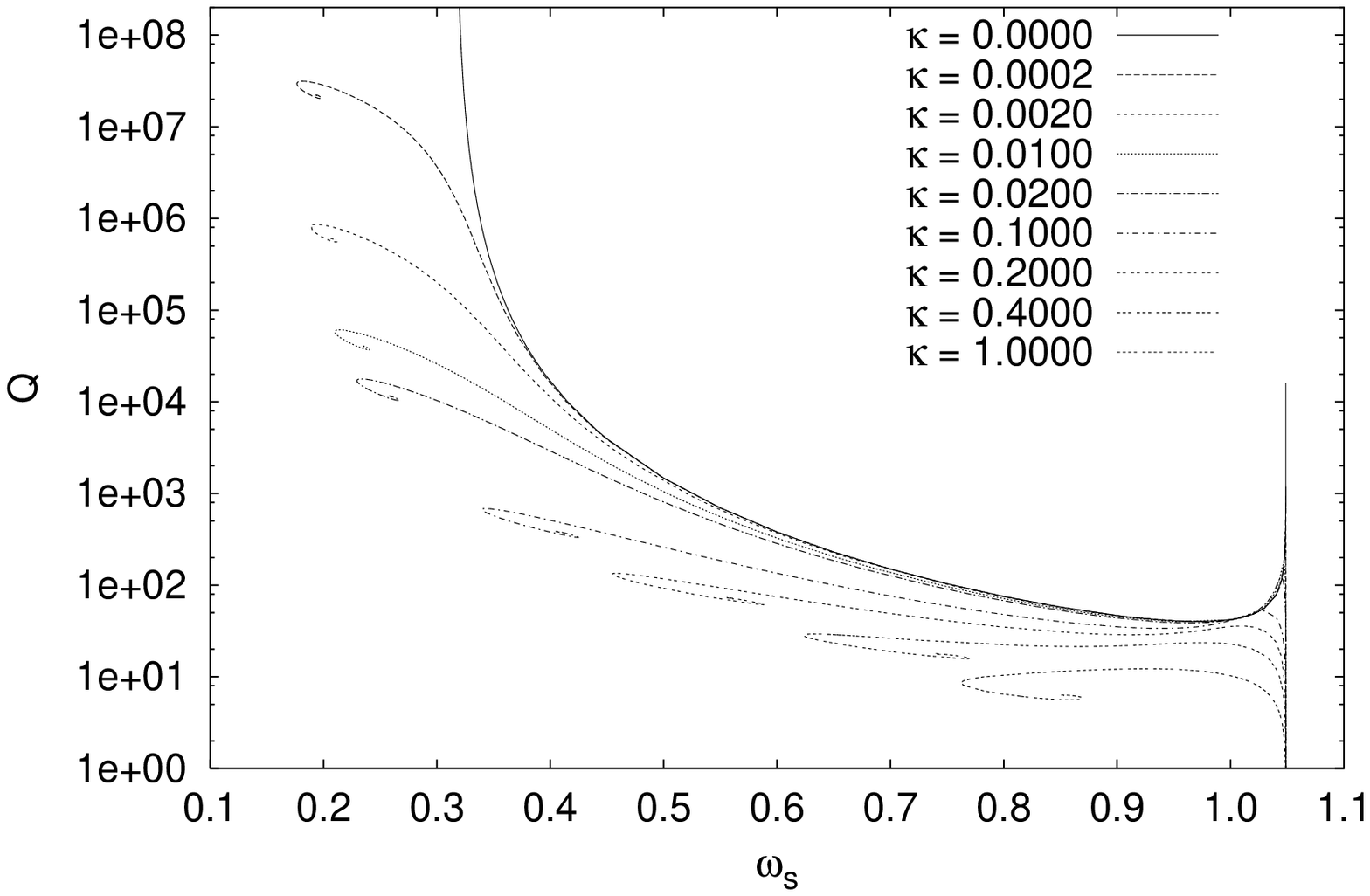}
\includegraphics[width=68mm,angle=0,keepaspectratio]{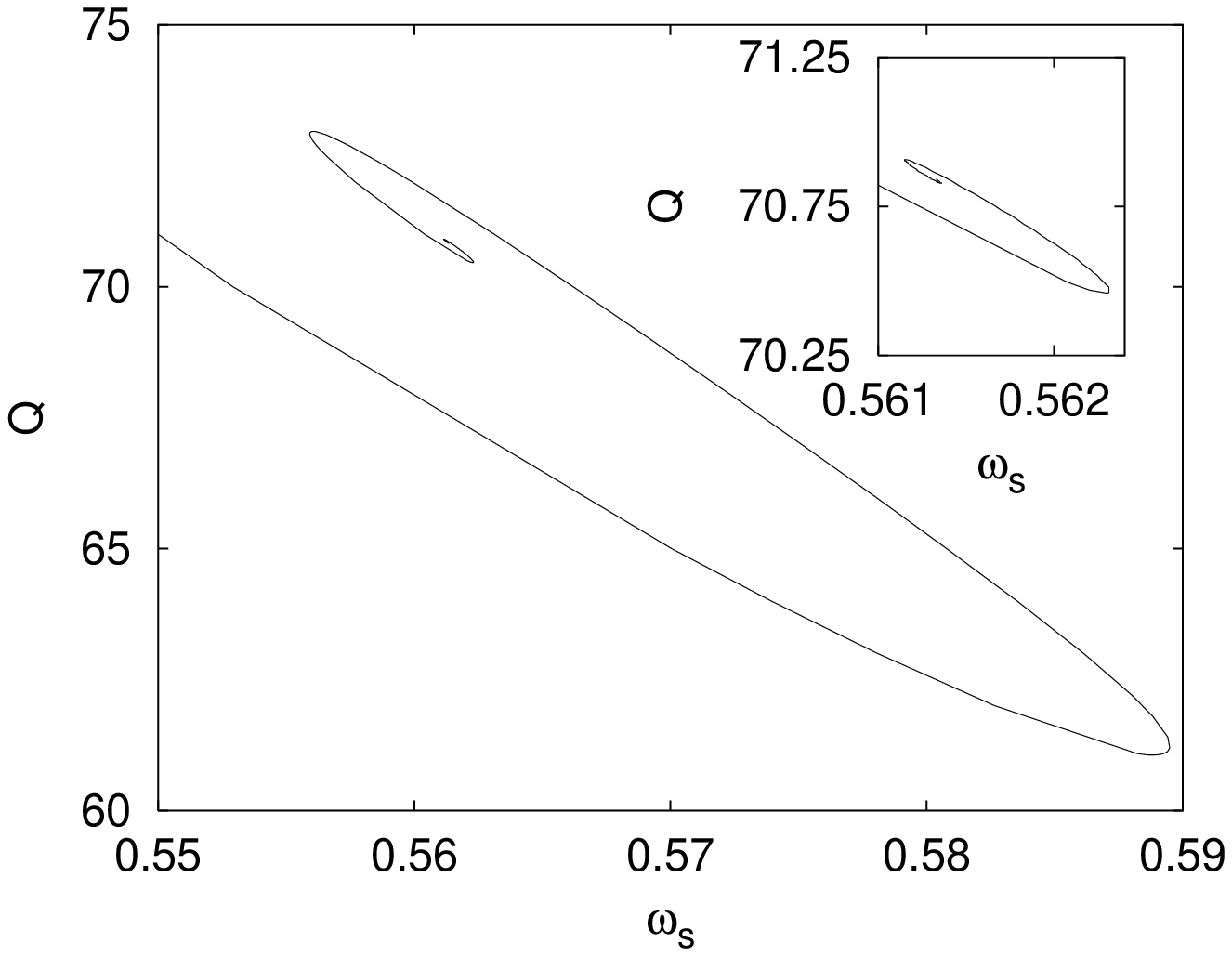}
}}}
\caption{
Left: The charge $Q$ versus the frequency $\omega_s$
for fundamental spherically symmetric boson stars ($k=0$) 
in the full range of existence 
for a sequence of values of the gravitational coupling constant $\kappa$.
Right:
The charge $Q$ versus the frequency $\omega_s$
in the frequency range of the spiral
for the gravitational coupling constant $\kappa=0.2$.
}
\label{qvsos_col}
\end{figure}

When the scalar field is coupled to gravity, boson stars arise
(see e.g.~\cite{jetzer,schunck2} for reviews).
We first demonstrate the effects of gravity for 
spherically symmetric boson stars in Fig.~\ref{qvsos_col},
where we exhibit the charge $Q$ versus the frequency $\omega_s$ 
for several values of the gravitational coupling constant $\kappa$.

For solutions in curved space the frequency $\omega_s$ 
is also bounded from above by
$\omega_{\rm max}$, Eq.~(\ref{cond1}),
ensuring an asymptotically exponential fall-off of the scalar field.
But, unlike for flat space solutions, 
the charge $Q$ does not diverge for boson stars,
when $\omega_s$ approaches $\omega_{\rm max}$.
Instead the charge tends to zero
in the limit $\omega_s \rightarrow \omega_{\rm max}$.

Also, for the smaller values of $\omega_s$ 
the solutions approach the limiting lower value 
$\omega_{\rm min}$, Eq.~(\ref{cond2}),
no longer monotonically.
Instead a spiral-like behaviour arises in the presence of gravity,
where the boson star solutions exhibit an inspiralling
towards a limiting solution \cite{lee-bs}.
The location and size of the spiral depend on the gravitational
coupling strength $\kappa$,
and can be well below $\omega_{\rm min}$
for small values of $\kappa$ 
(see Fig.~\ref{qvsos_col})
\cite{list}.

\begin{figure}[h!]
\parbox{\textwidth}
{\centerline{
\mbox{
\epsfysize=10.0cm
\includegraphics[width=70mm,angle=0,keepaspectratio]{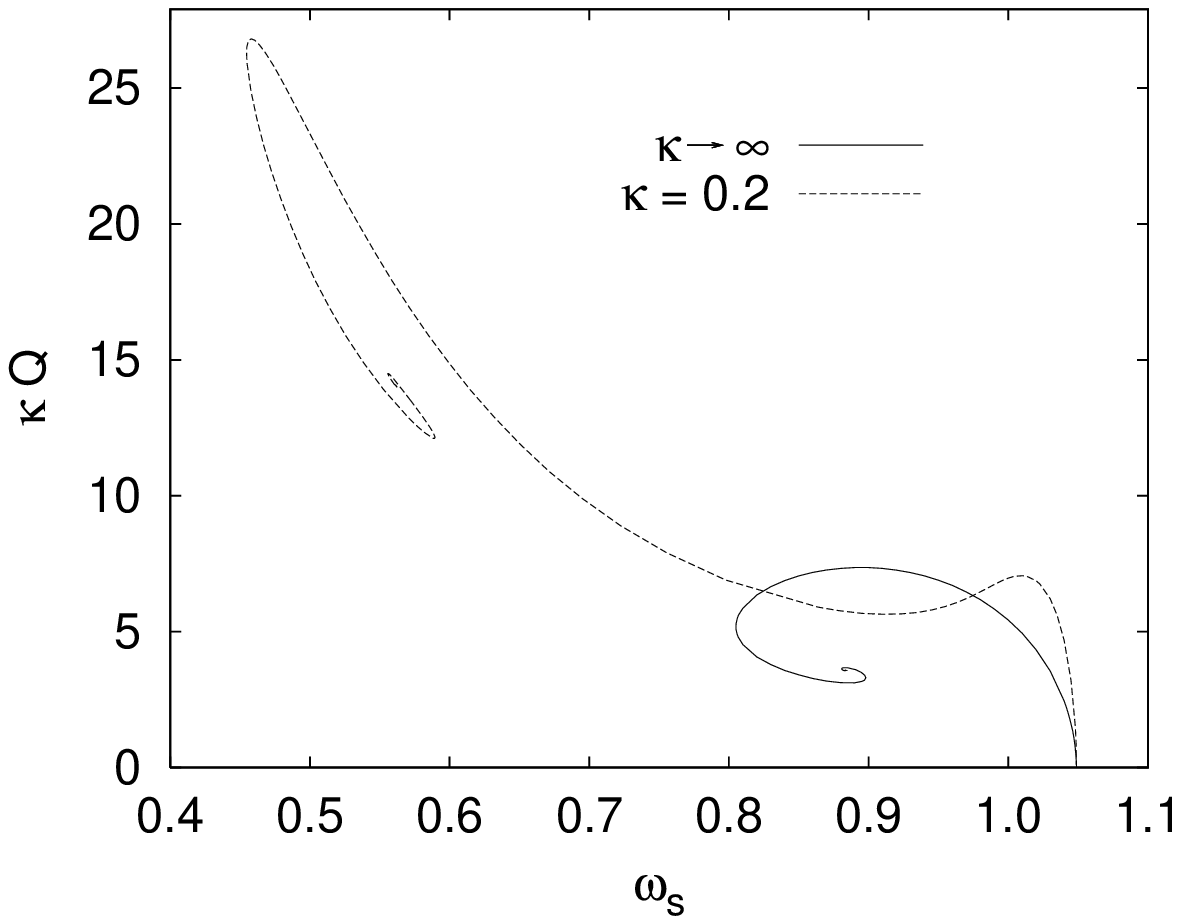}
\includegraphics[width=75mm,angle=0,keepaspectratio]{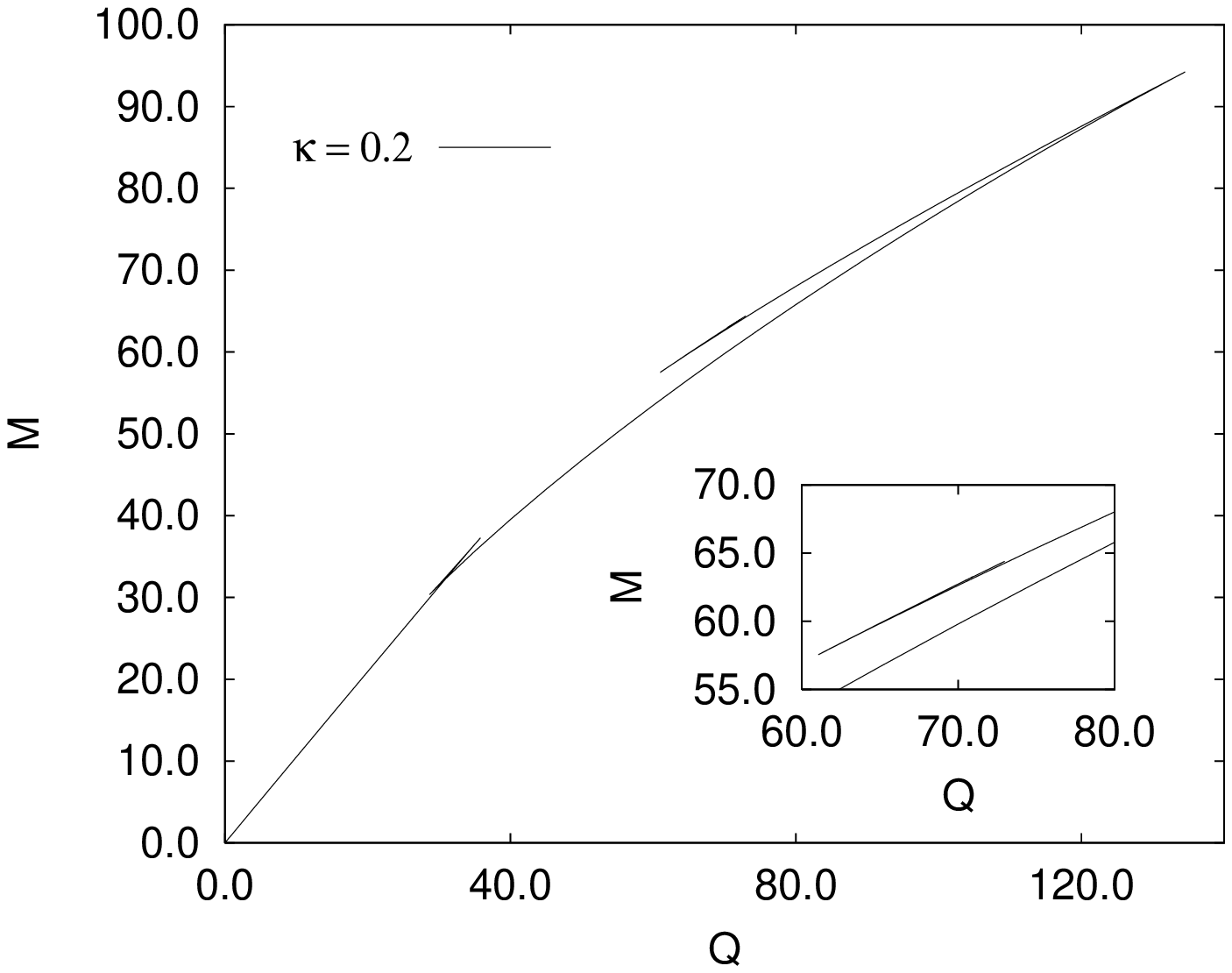}
}}}
\caption{
Left: The scaled charge $\kappa Q$
versus the frequency $\omega_s$
for fundamental spherically symmetric boson stars 
in the limit $\kappa \rightarrow \infty$.
For comparison the scaled charge $\kappa Q$
is also shown for $\kappa = 0.2$.
Right: The mass $M$ versus the charge $Q$
for fundamental spherically symmetric boson stars 
for the value of the gravitational coupling constant
$\kappa=0.2$
}
\label{mvsq_col}
\end{figure}

The limit $\kappa \rightarrow \infty$
is obtained by introducing the scaled
scalar field $\hat{\phi}(r)=\sqrt{\kappa} \phi(r)$ \cite{list}. 
Substituting $\phi(r)$ in the field equation and taking the
limit $\kappa \rightarrow \infty$, the terms non-linear 
in $\hat{\phi}(r)$ vanish. Similarly, in the Einstein equations 
the terms of higher than second order in $\hat{\phi}(r)$ vanish and
the dependence on $\kappa$ cancels.
The resulting set of differential equations is then identical to
the original one, except that $\kappa=1$ and 
$U(\hat{\phi})= \lambda b \hat{\phi}^2$,
reducing the potential to a mass term.
This also yields a spiral pattern for the
(scaled) charge $\hat{Q} = \kappa Q$, as seen in Fig.~\ref{mvsq_col}.

The mass $M$ has an analogous dependence on the frequency $\omega_s$
as the charge $Q$.
When the mass $M$ is considered as a function of the charge $Q$, however,
one now observes a new intricate cusp structure \cite{lee-bs,list}.
As gravity is weekly coupled,
the single cusp present in flat space
(associated with the minimal value $Q_{\rm min}$ of the charge)
is retained. But it is supplemented by a second cusp,
since in the presence of gravity the charge and the mass 
do not diverge for $\omega_s \rightarrow \omega_{\rm max}$,
but tend to zero in this limit.
For larger values of the gravitational coupling $\kappa$ 
these two cusps merge and disappear.
However, an additional set of cusps arises in the presence of gravity, 
which is due to the appearance of the spirals,
and therefore only present in curved space \cite{lee-bs,list}.
The cusp structure is illustrated in Fig.~\ref{mvsq_col}.

The classical stability of boson stars can be analyzed according
to catastrophe theory,
implying a change of classical stability at each cusp encountered
\cite{catastrophe}.
In the following we refer to the main branch of solutions 
as the branch between
the maximum of the charge and the mass 
(at a relatively small value of the frequency $\omega_s$)
and the 
local minimum of the charge and the mass
(at a relatively large value of the frequency $\omega_s$).
Note, that for the flat space solutions the minimum is a global minimum,
the charge and the mass assume their critical values here,
while for large gravitational coupling $\kappa$ the local minimum
has disappeared, and the relevant minimum corresponds to the global
minimum at $\omega_{\rm max}$.

Along this main branch the boson star solutions are classically stable.
We note though, that a small part of this main branch 
may become quantum mechanically unstable, 
when $M>m_{\rm B}Q$ close to the first cusp.
Since at each cusp terminating the main branch a negative mode is acquired,
the solutions on the adjacent branches become classically unstable.
In the spiral, the boson stars are expected to acquire
at each further cusp an additional negative mode,
making the solutions increasingly unstable.
The physically most relevant set of solutions
should thus correspond 
to the stable boson star solutions on the main branch.

\boldmath
\subsection{Rotating Boson Stars}\label{c4}
\unboldmath

We now turn to rotating boson stars,
which emerge from rotating $Q$-balls, 
when the gravitational coupling constant is increased from zero.
We first recall the previous results
on rotating positive parity boson stars with $n^P=1^+$ 
\cite{schunck,japan,list}
and present new results for $n^P=2^+$.
We then turn to rotating negative parity boson stars.
For both cases we demonstrate and analyze the occurrence of ergoregions.
We here restrict our analysis to fundamental rotating boson stars.

\subsubsection{Positive parity boson stars}

\begin{figure}[h!]
\parbox{\textwidth}
{\centerline{
\mbox{
\epsfysize=10.0cm
\includegraphics[width=70mm,angle=0,keepaspectratio]{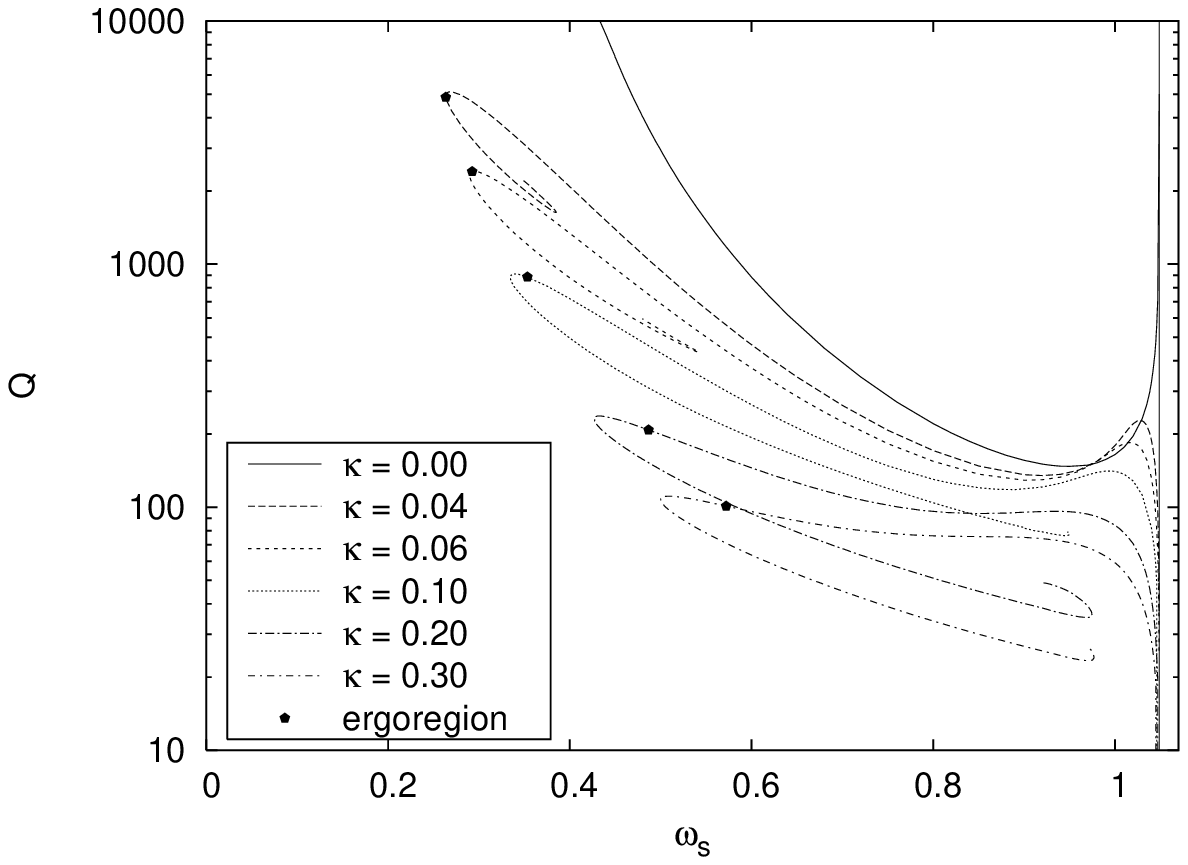}
\includegraphics[width=70mm,angle=0,keepaspectratio]{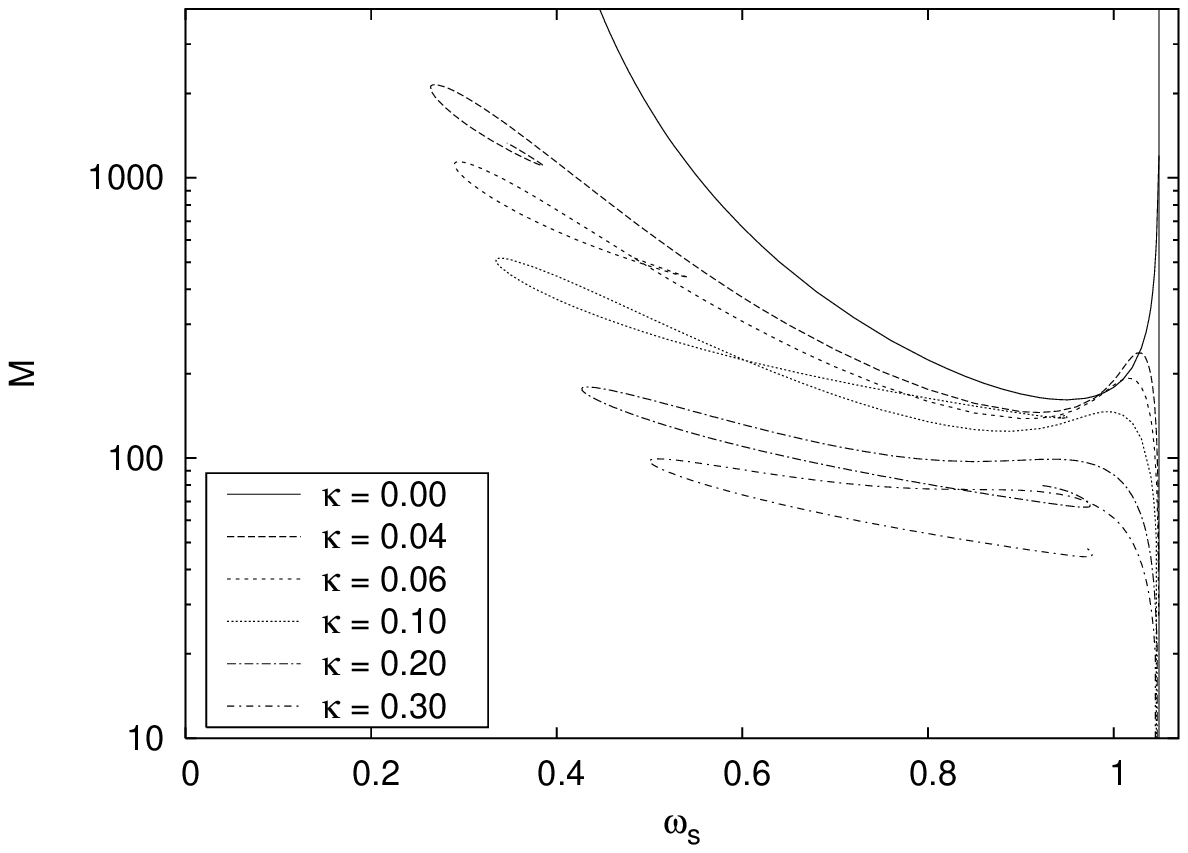}
}}}
\caption{
Left:
The charge $Q$ versus the frequency $\omega_s$
for rotating $1^+$ boson stars
for several values of the gravitational coupling constant
$\kappa$. Also shown are the limiting flat space values.
The dots indicate the onset of ergoregions.
Right:
The mass $M$ versus the frequency $\omega_s$
for the same set of solutions.
}
\label{Qlimrot}
\end{figure}

We begin the discussion by recalling the main results on
boson stars with $n^P=1^+$, obtained previously \cite{list}.
For that purpose we exhibit in Fig.~\ref{Qlimrot} 
the charge $Q$ of $1^+$ boson stars versus the frequency $\omega_s$
for several values of the gravitational coupling constant $\kappa$,
including the flat space limit.

Again, the frequency $\omega_s$ of the solutions
is bounded from above by
$\omega_{\rm max}$, Eq.~(\ref{cond1}),
ensuring an asymptotically exponential fall-off
of the scalar field.
Also, the charge of the rotating boson stars tends to zero,
when $\omega_{\rm max}$ is approached.
Likewise, $1^+$ boson stars 
exist only in a finite interval of the frequency $\omega_s$, 
where the lower bound increases with increasing $\kappa$.

Furthermore, we observe the presence of spirals also
for these rotating boson stars,
i.e., the $1^+$ boson stars exhibit an inspiralling
towards a limiting solution just like the non-rotating boson stars.
For small values of the coupling constant $\kappa$ the spirals are
located in the region of small frequency $\omega_s$,
for larger coupling $\kappa$ the spirals shift towards larger frequencies.
A good and rather complete determination of the spirals is, however,
numerically often extremely difficult \cite{list}.

\begin{figure}[h!]
\parbox{\textwidth}
{\centerline{
\mbox{
\epsfysize=10.0cm
\includegraphics[width=70mm,angle=0,keepaspectratio]{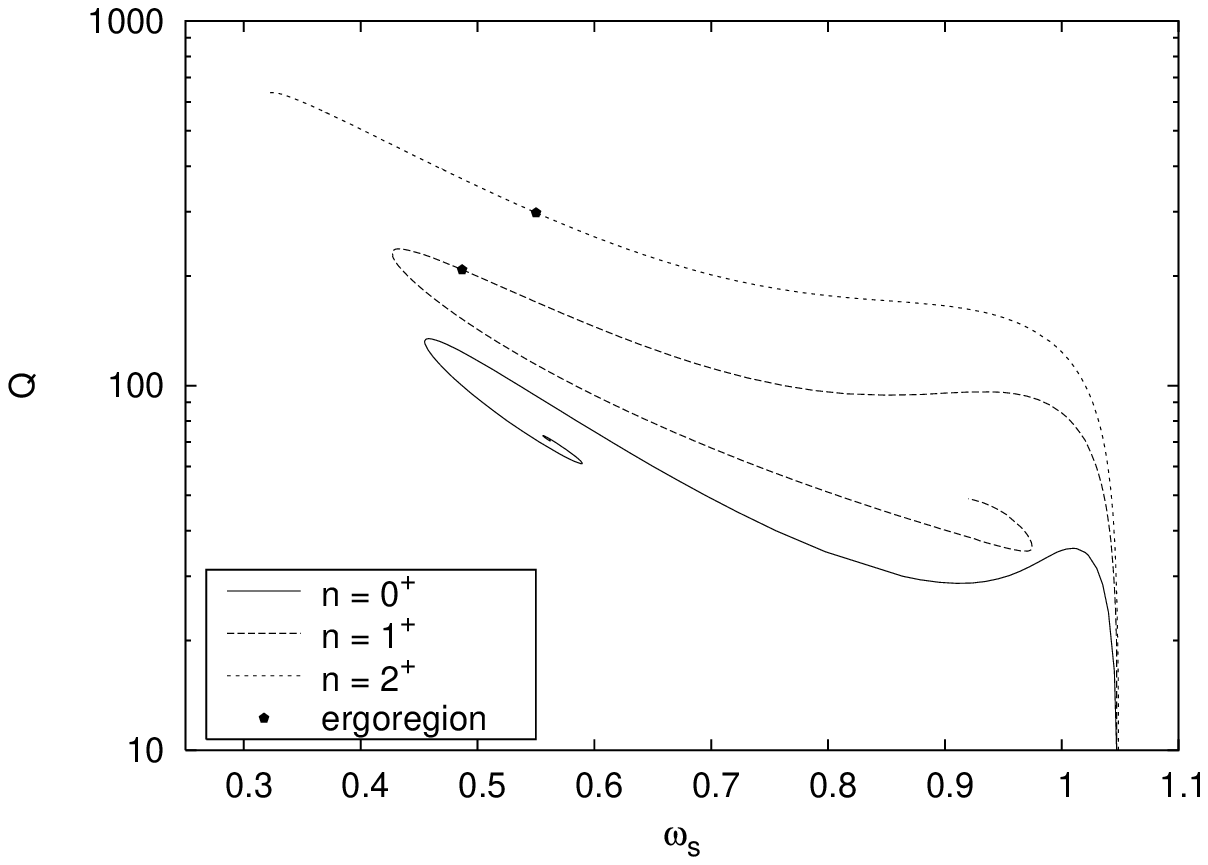}
\includegraphics[width=70mm,angle=0,keepaspectratio]{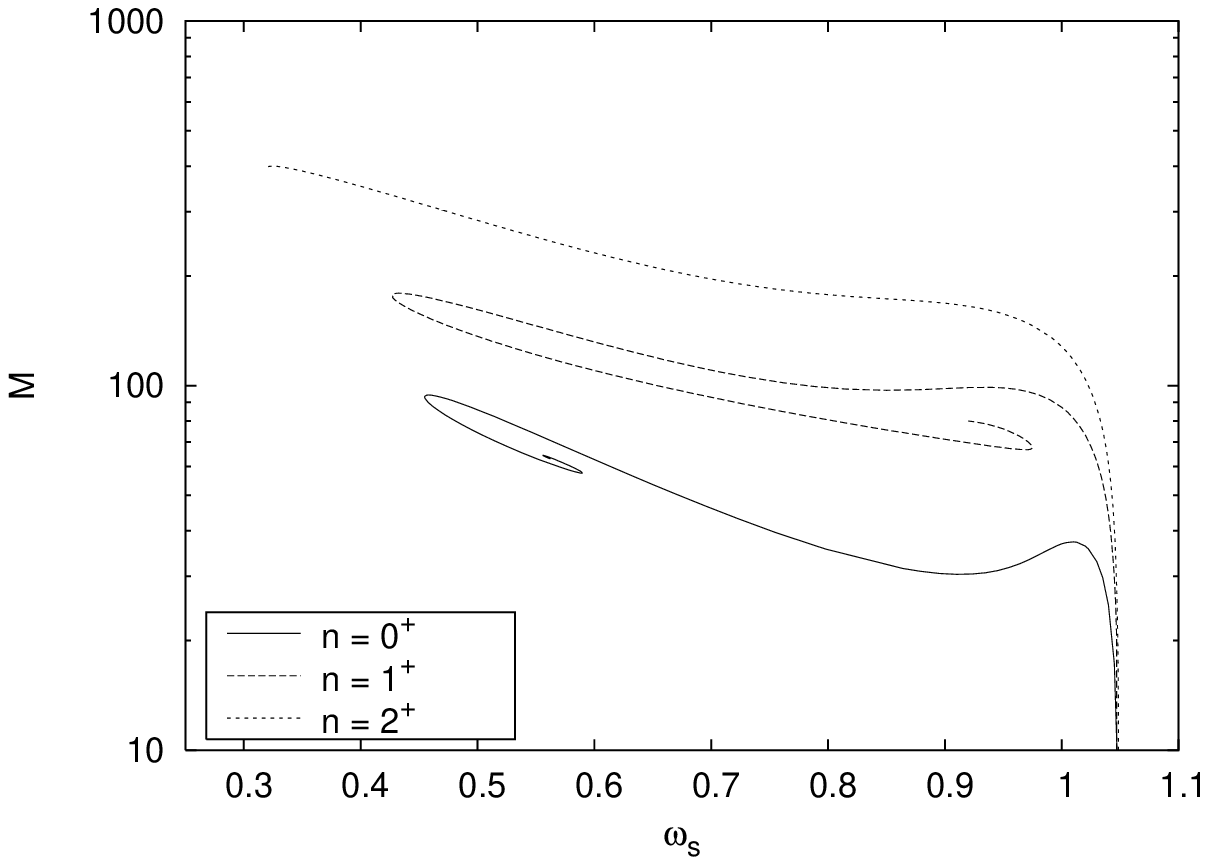}
}}}
\caption{
The charge $Q$ (left column) and the mass $M$ (right column)
versus the frequency $\omega_s$
for boson stars with $n^P=0^+$, $1^+$, $2^+$
for the gravitational coupling $\kappa=0.2$.
}
\label{Qrotbs}
\end{figure}

The mass $M$ of the $1^+$ boson stars
has an analogous dependence on the frequency $\omega_s$
as the charge $Q$, as seen in Fig.~\ref{Qrotbs}.
When the mass $M$ of these rotating boson stars
is considered as a function of the charge $Q$,
we observe an analogous cusp structure as for the non-rotating boson stars
\cite{list}.
In particular, for a given gravitational coupling $\kappa$,
a main branch of rotating boson star solutions
is present, and located between
the maximum of the charge and the mass
(at a relatively small value of the frequency $\omega_s$)
and the (for small gravitational coupling only local)
minimum of the charge and the mass
(which is not part of the spiral).
Following arguments from catastrophe theory,
the boson star solutions are again 
expected to be classically stable along this main branch.

\begin{figure}[h!]
\parbox{\textwidth}
{\centerline{
\mbox{
\epsfysize=10.0cm
\includegraphics[width=70mm,angle=0,keepaspectratio]{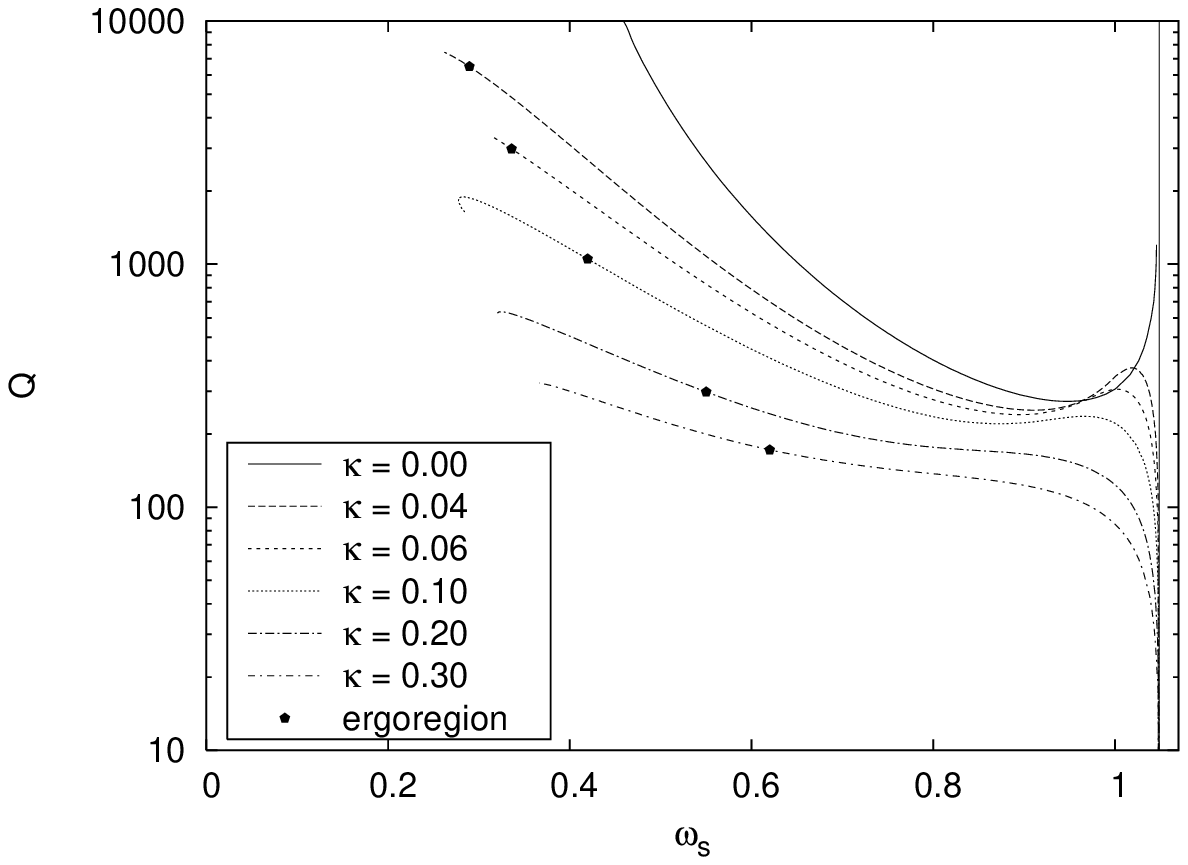}
\includegraphics[width=70mm,angle=0,keepaspectratio]{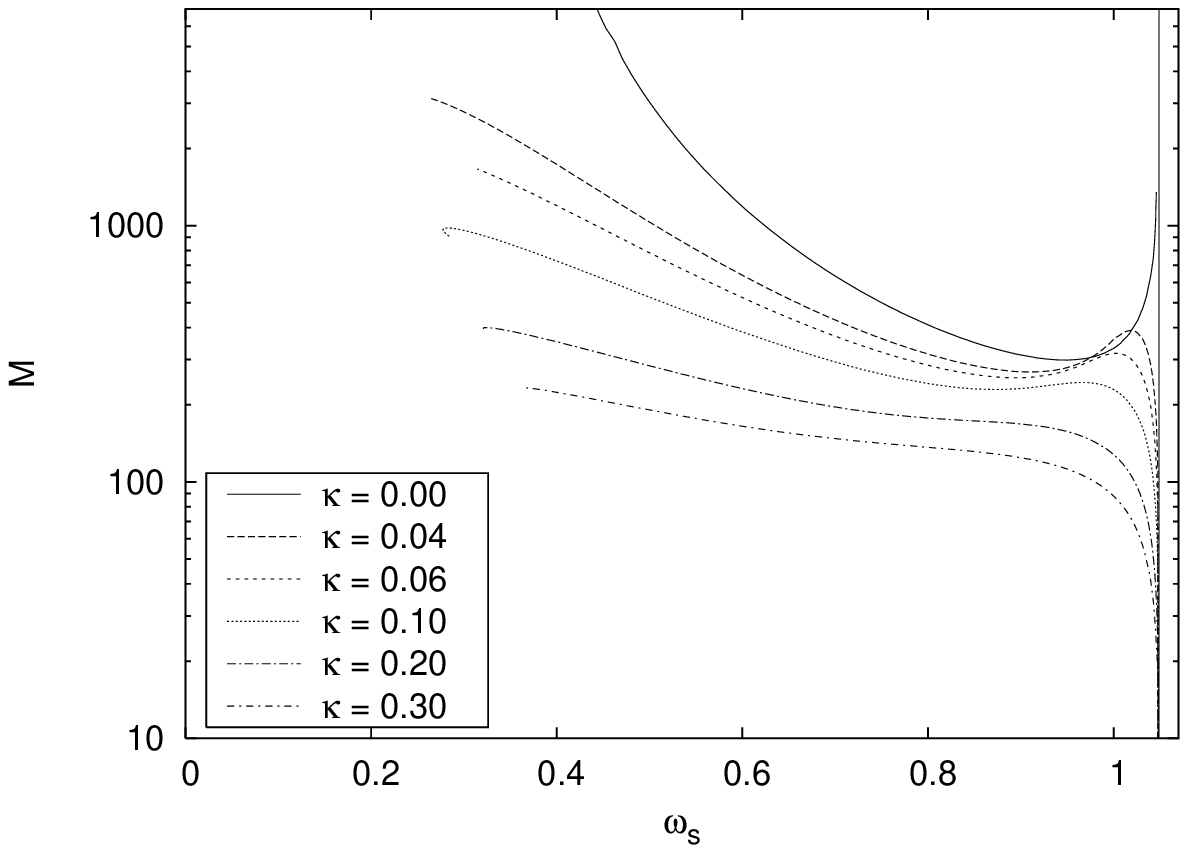}
}}}
\caption{
Left:
The charge $Q$ versus the frequency $\omega_s$
for rotating $2^+$ boson stars
for several values of the gravitational coupling constant
$\kappa$. Also shown are the limiting flat space values.
The dots indicate the onset of ergoregions.
Right:
The mass $M$ versus the frequency $\omega_s$
for the same set of solutions.
}
\label{Qlimrot2}
\end{figure}

Let us now turn to positive parity boson stars with higher
quantum number $n$ and thus greater angular momentum.
We exhibit in Fig.~\ref{Qlimrot2}
the charge $Q$ and the mass $M$
of $2^+$ boson stars versus the frequency $\omega_s$
for several values of the gravitational coupling constant $\kappa$,
including the flat space limit.
Because these calculations are numerically very involved
and time-consuming, we have concentrated on the main branches
of these $2^+$ boson stars, omitting the numerically even more involved
spirals.
Clearly, the $2^+$ solutions show a completely analogous
$\kappa$- and frequency dependence (in the regions studied)
as the $1^+$ solutions.

To further demonstrate the dependence of the positive parity boson stars 
on the rotational quantum number $n$, we exhibit in Fig.~\ref{Qrotbs}
the charge $Q$ and the mass $M$ as functions of the
frequency $\omega_s$ for boson stars with 
quantum numbers $n^P=0^+$, $1^+$, $2^+$  
and gravitational coupling $\kappa=0.2$.

Considering the spatial structure of the solutions,
the scalar field $\f$ typically gives rise to
a torus-like energy density $T_{tt}$ 
for the rotating $1^+$ and $2^+$ boson stars,
just as for the rotating $n^+$ $Q$-balls.
Again, the torus-like shape becomes apparent by
considering surfaces of constant energy density
of e.g.~half the respective maximal value of the energy density.
As in the flat space solutions, the maximum of the energy density
of the boson stars increases and shifts towards
larger values of the coordinate $\rho=r \sin \theta$,
when $n$ and thus the angular momentum increases.
Thus in terms of the energy density tori this again manifests as an
increase of their radii in the equatorial plane with increasing $n$.

Rotating objects may possess ergoregions.
But unlike the ergoregions of black holes, the ergoregions of 
regular objects like boson stars
would signal the presence of an instability
\cite{Cardoso,ergo-paper1,ergo-paper2,ergo-paper3}.
This instability can be traced back to superradiant scattering,
and its possible presence was put forward by Cardoso et al.~\cite{Cardoso}
as an argument against various black hole doubles
as potential horizonless candidates for compact dark astrophysical objects.

Therefore we now analyze the occurrence of ergoregions for
rotating positive parity boson stars.
The ergosurface is defined by the condition
\begin{equation}
g_{tt}= 0 = -f + \frac{l}{f} \sin^2 \theta \omega^2 \ 
\label{ergo-c}
\end{equation}
in the metric parametrization Eq.~(\ref{ansatzg}).
The ergoregion resides inside this ergosurface.
The presence of an ergosurface thus implies instability 
of the respective boson stars \cite{Cardoso}.

Examining the condition $g_{tt}= 0$ for the sets of boson star solutions,
we indeed observe that rotating boson stars possess ergoregions
in a large part of their domain of existence.
To discuss these ergoregions, let us recall Figs.~\ref{Qlimrot} and 
\ref{Qlimrot2}, where the charge $Q$ is exhibited versus the
frequency $\omega_s$ for several sets of boson stars solutions with
$n^P=1^+$ and $2^+$, respectively.
The dots in these figures indicate the onset of ergoregions
and thus ergoregion related instability 
at a corresponding critical frequency $\omega_s$.
The solutions to the right of the dots do not possess an ergoregion,
whereas the solutions to the left of the dots as well as in the spirals
do possess an ergoregion.

The onset of ergoregions thus almost always occurs on the main branch
of boson star solutions, supposed to be classically stable.
The presence of ergoregions then tends to diminish the physically
relevant - since stable - set of solutions.
As seen from the figures, the diminishment is greater
for larger gravitational coupling and larger rotational quantum number $n$,
and thus greater angular momentum.
However, there always remains a part of the main branch
of classically stable boson star solutions, 
not suffering from an ergoregion instability.

\begin{figure}[h!]
\parbox{\textwidth}
{\centerline{
\mbox{
\epsfysize=10.0cm
\includegraphics[width=75mm,angle=0,keepaspectratio]{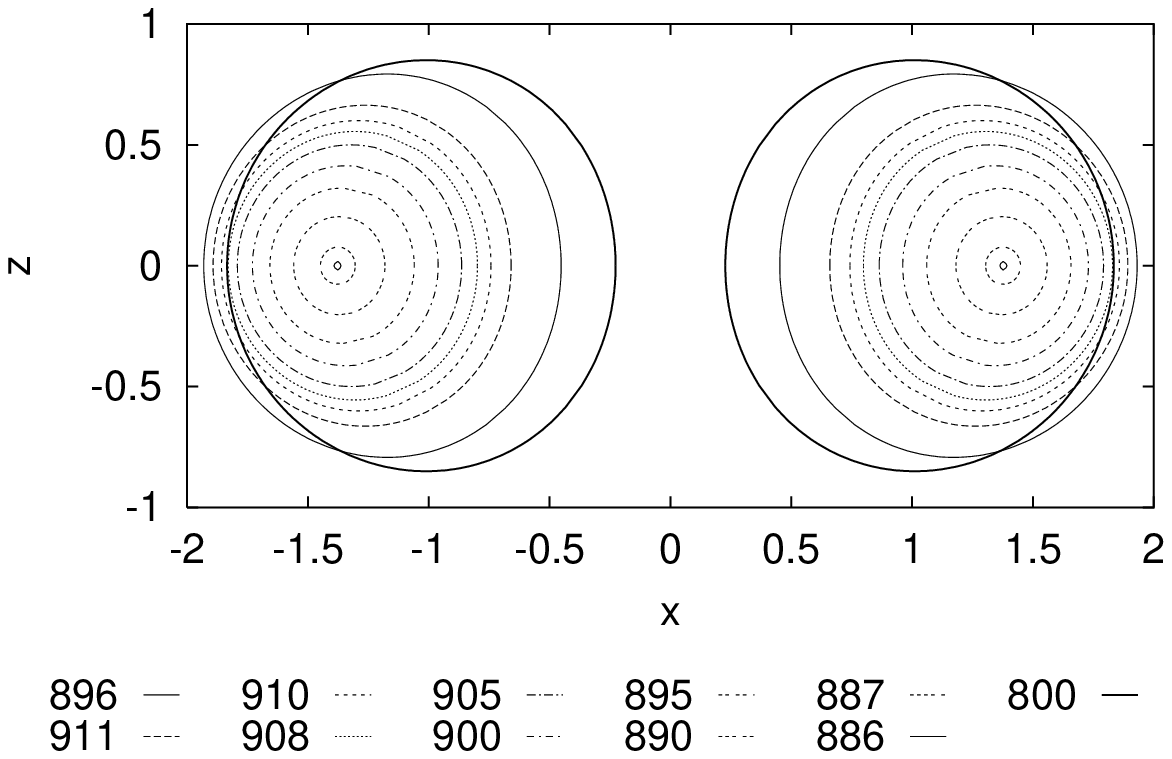}
\epsfysize=10.0cm
\includegraphics[width=75mm,angle=0,keepaspectratio]{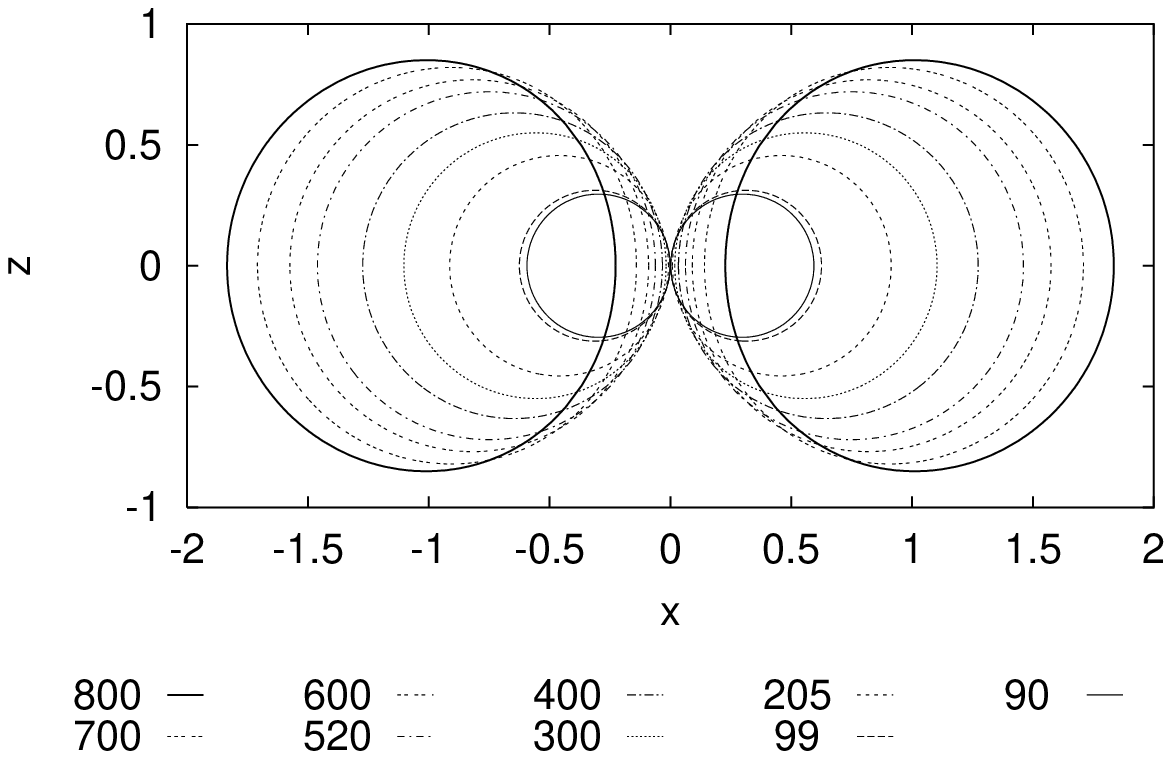}
}}}
{\centerline{
\mbox{
\epsfysize=10.0cm
\includegraphics[width=75mm,angle=0,keepaspectratio]{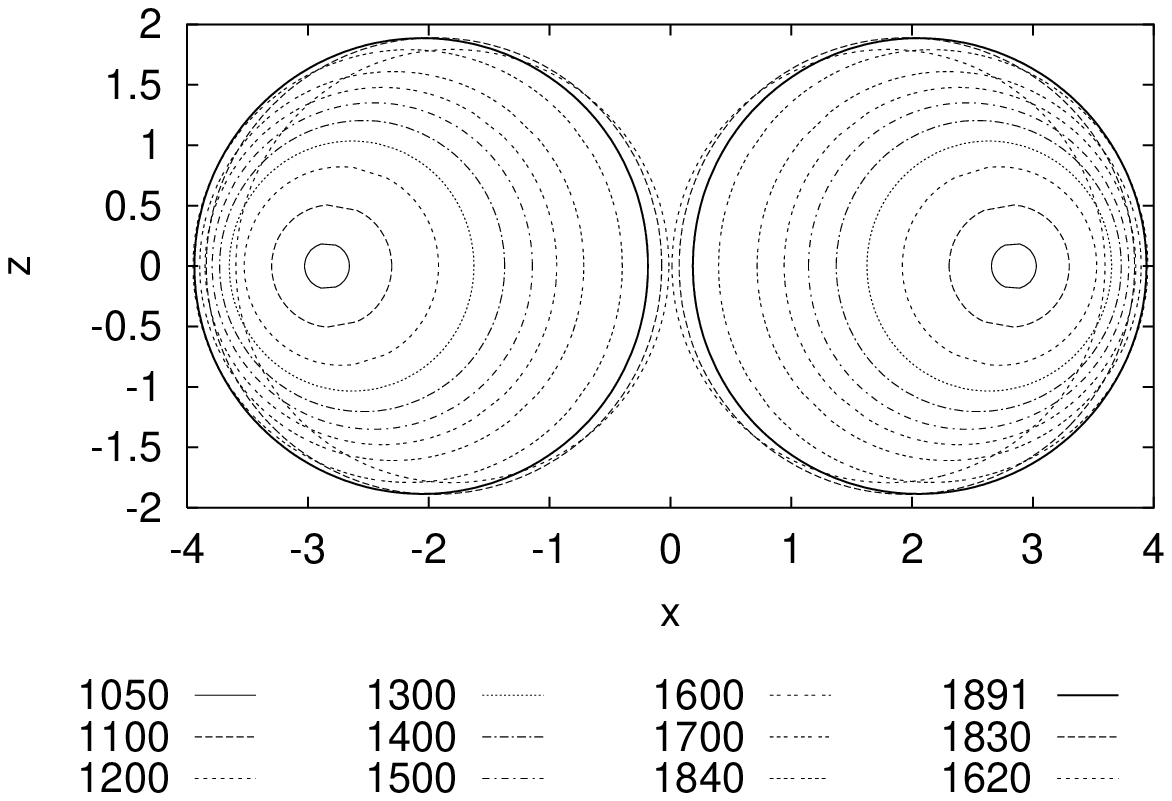}
\epsfysize=10.0cm
\includegraphics[width=75mm,angle=0,keepaspectratio]{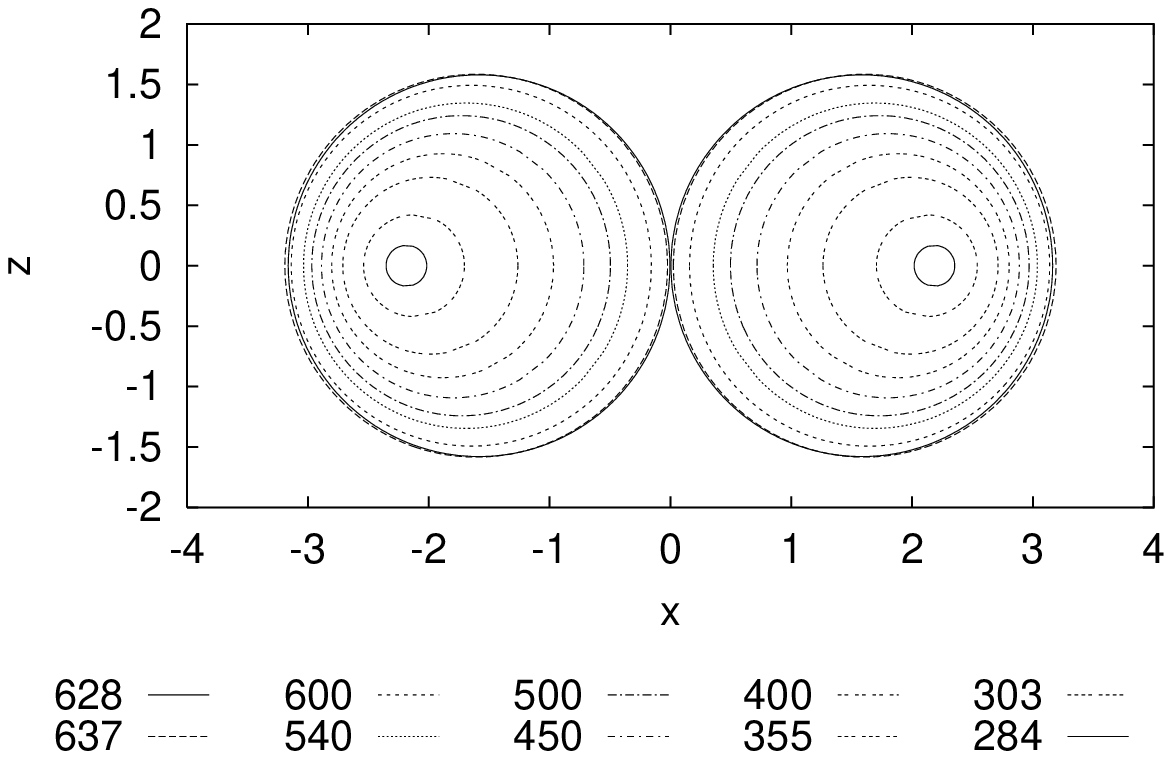}
}}}
\caption{
Location and shape of the axially symmetric ergoregions 
of boson stars exhibited via cross-sections with the $xz$-plane:
$1^+$ boson stars at $\kappa=0.1$ (upper row)
and $2^+$ boson stars (lower row) 
at $\kappa=0.1$ (left) and at $\kappa=0.2$ (right).
The numbers in the legend indicate the respective values of the charge $Q$.
}
\label{ergorot}
\end{figure}

Let us now discuss the emergence and structure of ergoregions
in more detail for these boson stars.
For that purpose we exhibit the ergoregions
of $1^+$ boson stars at $\kappa=0.1$
and $2^+$ boson stars at $\kappa=0.1$ and $0.2$
in Fig.~\ref{ergorot}.

Beginning with the $1^+$ boson stars, we observe, that
at the critical frequency (where $Q\approx 986$ for $\kappa=0.1$)
the condition (\ref{ergo-c}) 
is satisfied precisely on a circle in the equatorial plane.
In the (upper left) figure, only the two points of intersection of this
circle with the $xz$-plane are seen.
As $Q$ is increased (i.e., $\omega_s$ is decreased)
the circle turns into a torus-like surface.
In the figure,
its intersection with the $xz$-plane is then seen as two small circles
($Q=911$).
From this approximately maximal value of the charge,
the set of solutions now enters the first part of the spiral,
where stability is lost anyway.

As the solutions evolve into the spiral the charge decreases.
But the ergoregion continues to grow in size, while it remains
(approximately) centered at the location, where it first arose.
The ergoregion then tends towards its largest size
($Q \approx 800$), while the center of the torus of the ergoregion
starts to shift inwards.
Interestingly, the ergoregion is not only topologically a torus, 
but also geometrically almost a torus.

As seen in the (upper right) figure, 
with further decreasing charge $Q$, 
the ergoregion then starts to decrease in size again,
when the solutions evolve deeper into the spiral.
At the same time, the center of the torus of the ergoregion
continues to move further inwards.
Thus the innermost part of the ergosurface gets increasingly
closer to the origin, but never touches it.

The ergoregions of the set of boson stars at different values
of the gravitational coupling constant $\kappa$
exhibit an analogous behaviour.
Likewise,
the ergoregions of the sets of boson stars at higher $n$
exhibit an analogous behaviour in the regions studied.
As seen in the (lower part of the) figure, the shift of the centers of the
tori of the ergoregion arises faster for $n=2$
than for $n=1$, yielding
almost touching ergosurfaces already in the vicinity
of the maximal size of the ergosurfaces.

\subsubsection{Negative parity boson stars}

\begin{figure}[h!]
\parbox{\textwidth}
{\centerline{
\mbox{
\epsfysize=10.0cm
\includegraphics[width=70mm,angle=0,keepaspectratio]{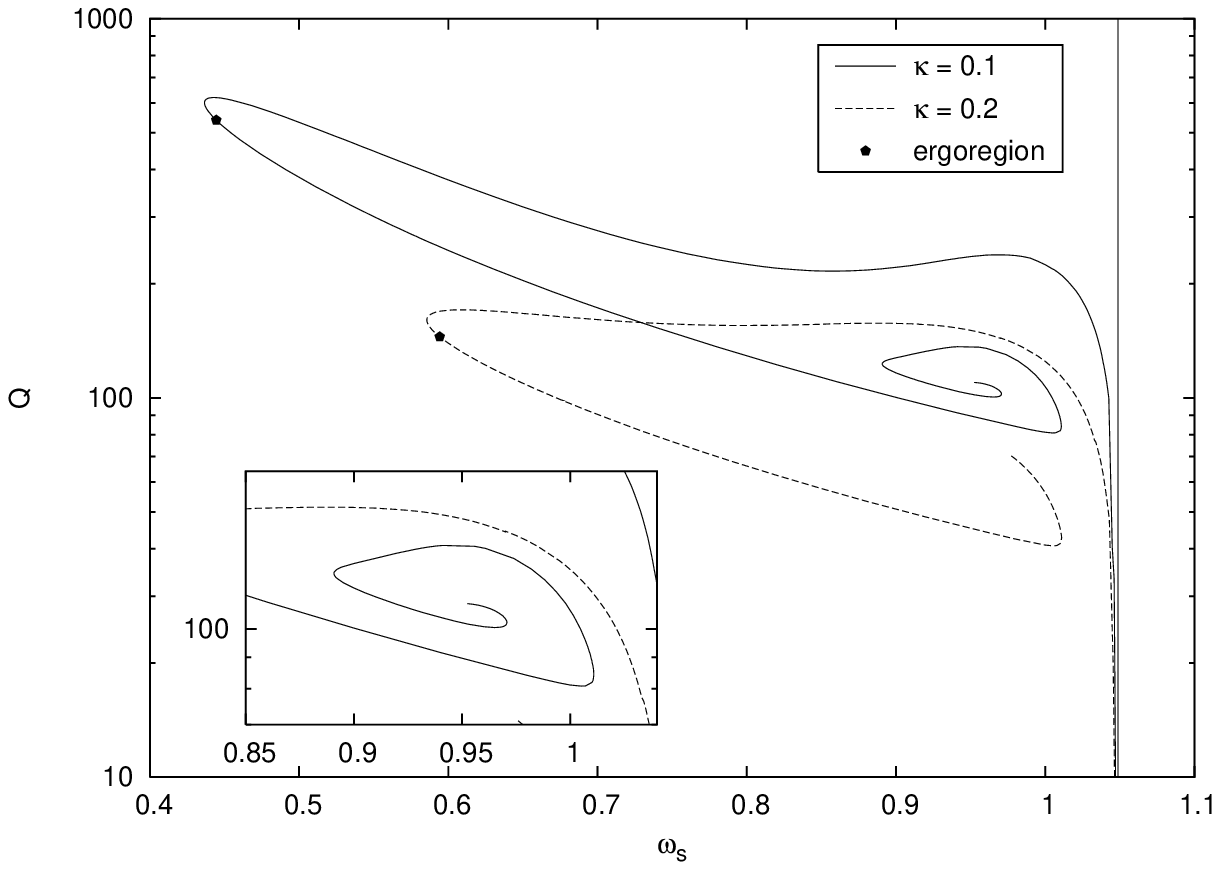}
\includegraphics[width=70mm,angle=0,keepaspectratio]{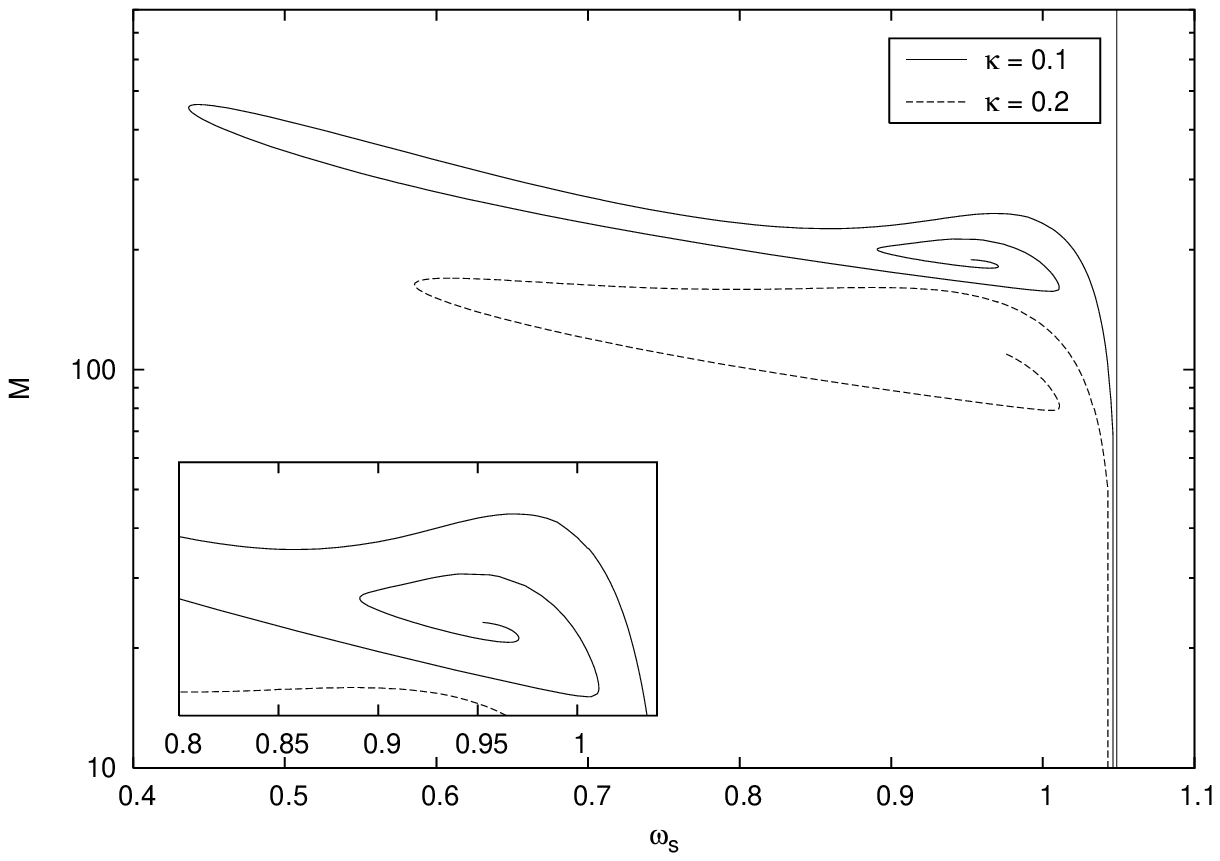}
}}}
\caption{
The charge $Q$ (left) and the mass $M$ (right)
versus the frequency $\omega_s$
for $1^-$ boson stars 
for gravitational coupling $\kappa=0.1$ and $0.2$.
The dots indicate the onset of ergoregions.
}
\label{Qrotbs1-}
\end{figure}

For boson stars with negative parity
we observe analogous features as for positive parity
boson stars. This is illustrated in the following.
We first consider $1^-$ boson stars
for gravitational coupling $\kappa=0.1$ and $0.2$.
The dependence of their charge $Q$ and mass $M$ 
on the frequency $\omega_s$ is exhibited
in Fig.~\ref{Qrotbs1-}. For $\kappa=0.1$ a large part of 
the spiral could be obtained numerically.

\begin{figure}[h!]
\parbox{\textwidth}
{\centerline{
\mbox{
\epsfysize=10.0cm
\includegraphics[width=70mm,angle=0,keepaspectratio]{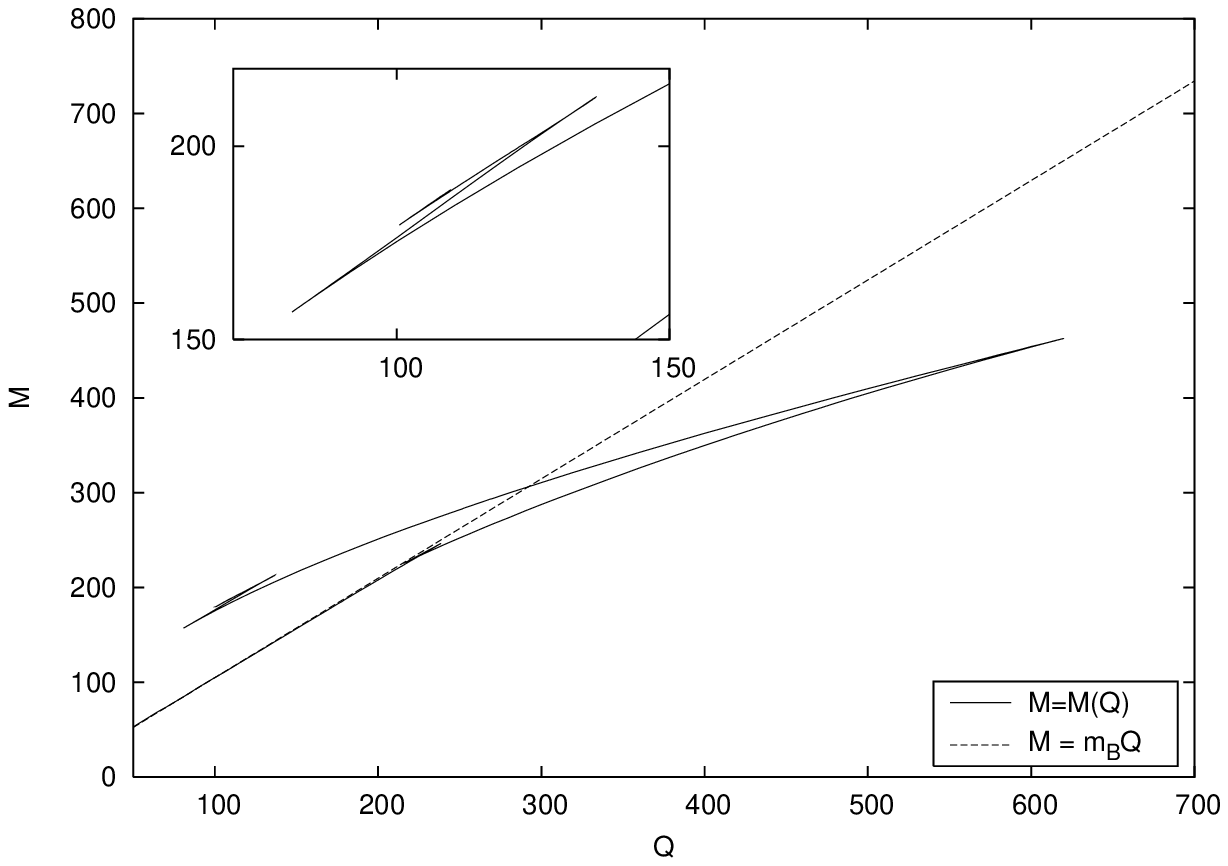}
\includegraphics[width=70mm,angle=0,keepaspectratio]{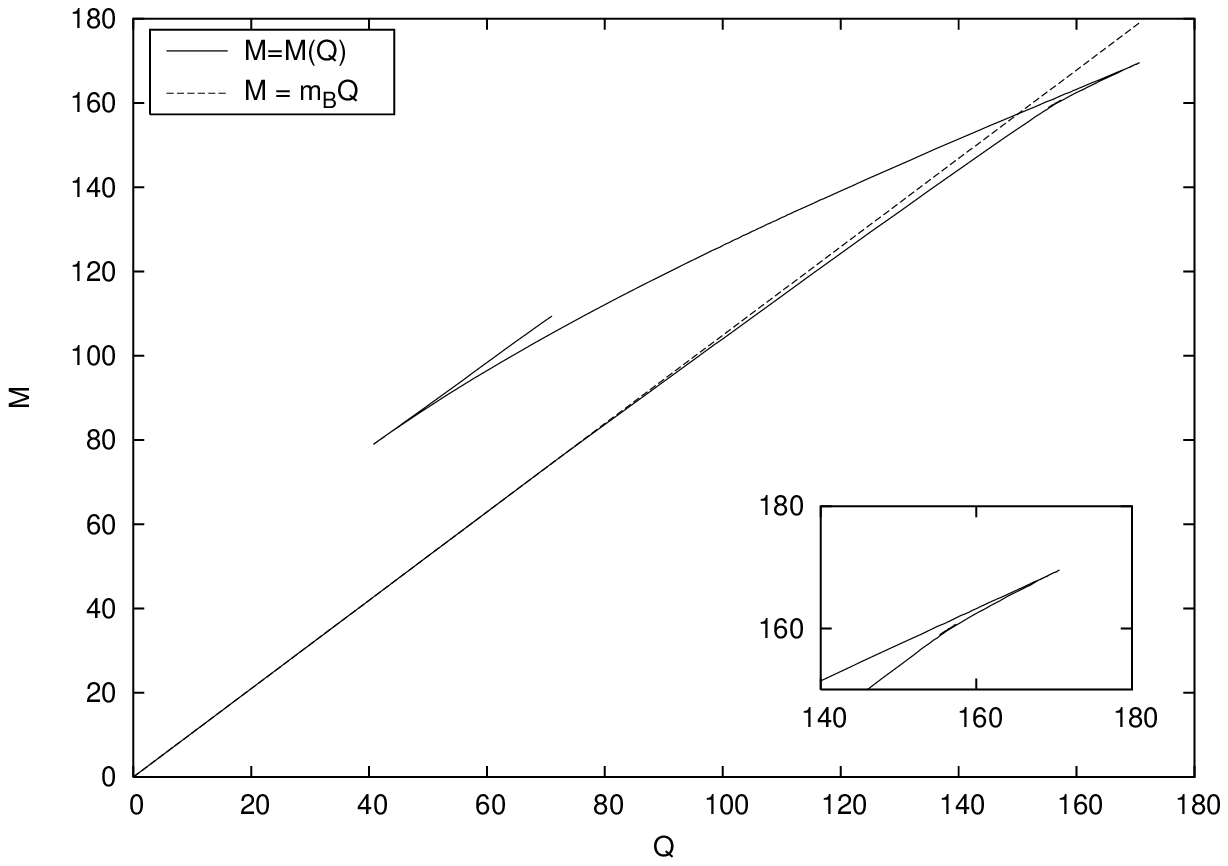}
}}}
\caption{
The mass $M$ versus the charge $Q$
for $1^-$ boson stars 
for gravitational coupling $\kappa=0.1$ (left) and $0.2$ (right),
together with the mass for $Q$ free bosons, $M=m_{\rm B}Q$.
}
\label{m_rot-}
\end{figure}

The mass $M$ of negative parity boson stars
exhibits the same characteristic cusp structure
as the mass of positive parity boson stars,
when considered as a function of the charge $Q$ \cite{list}.
This is seen in Fig.~\ref{m_rot-}
for $1^-$ boson stars
at gravitational coupling $\kappa=0.1$ and $0.2$.

\begin{figure}[h!]
\parbox{\textwidth}
{\centerline{
\mbox{
\epsfysize=10.0cm
\includegraphics[width=70mm,angle=0,keepaspectratio]{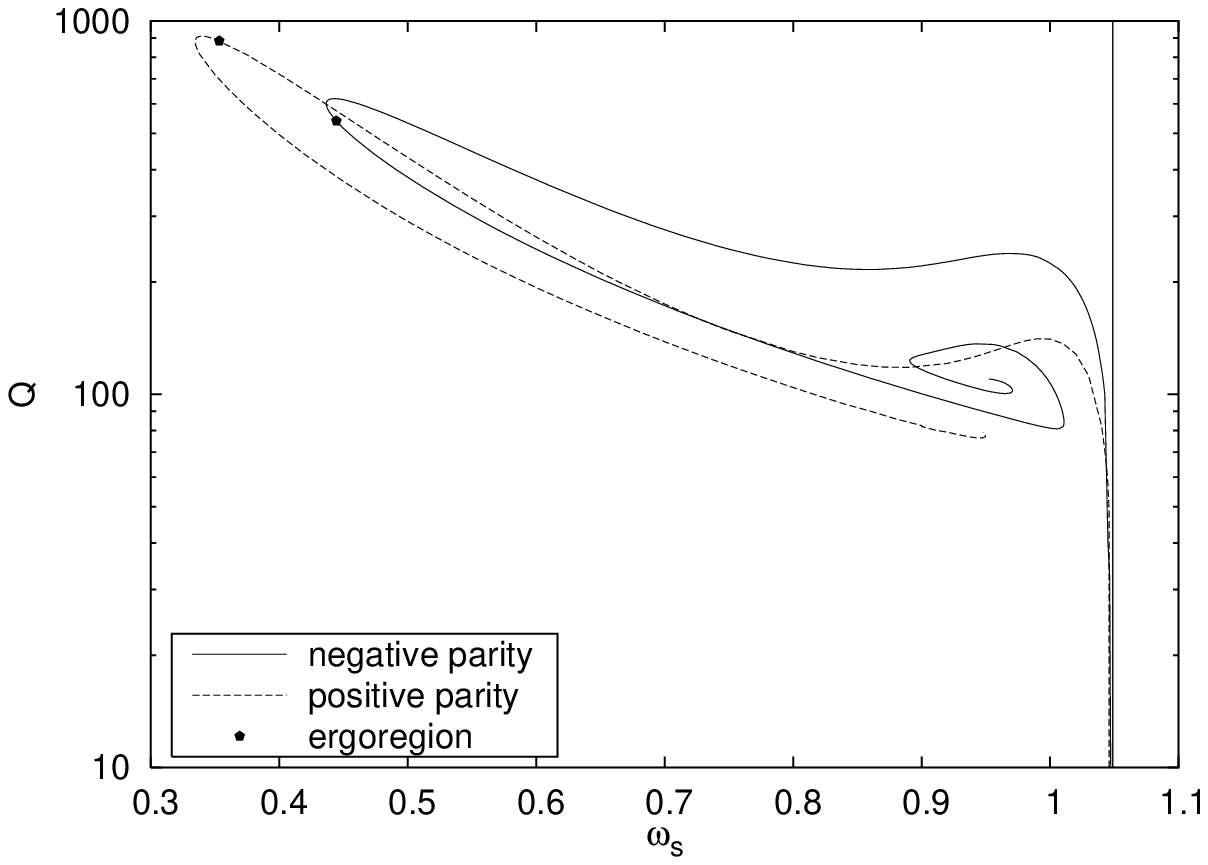}
\includegraphics[width=70mm,angle=0,keepaspectratio]{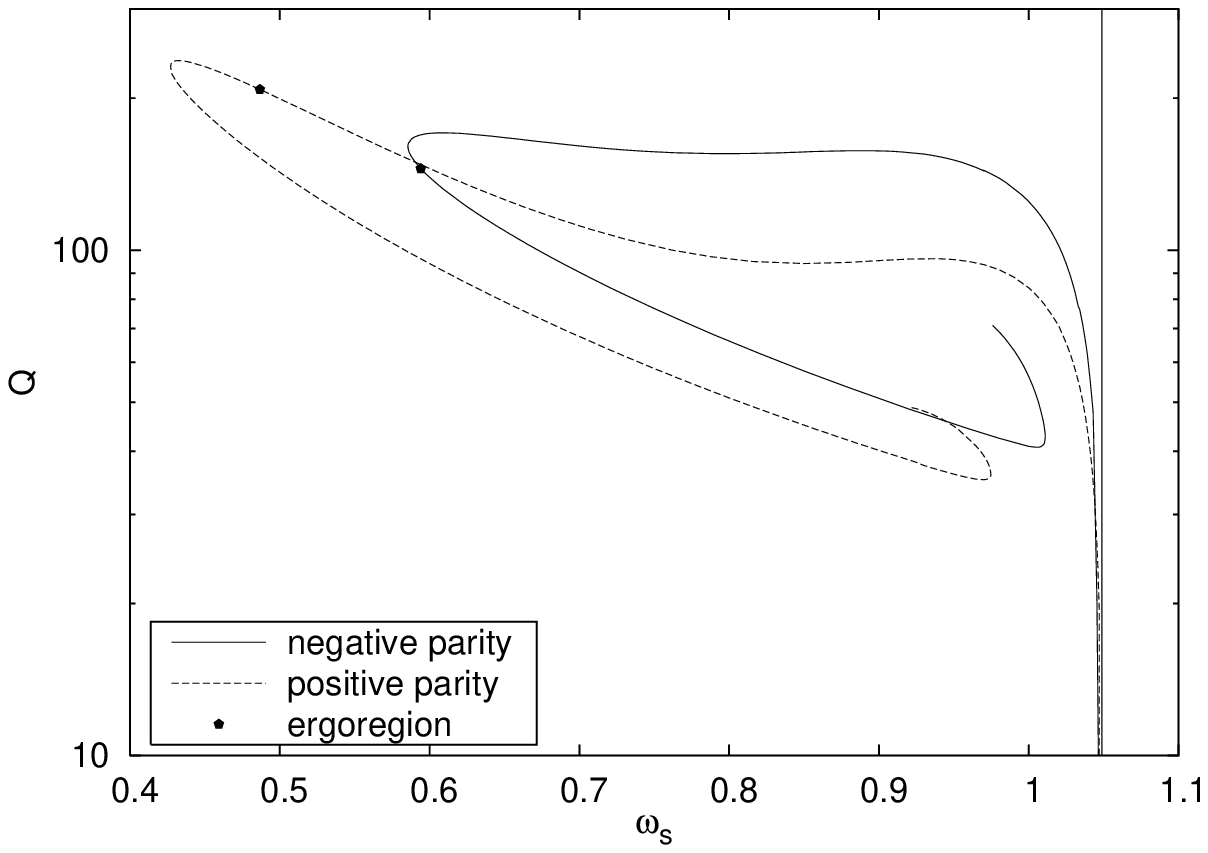}
}}}
\caption{
The charge $Q$ 
versus the frequency $\omega_s$
for $1^-$ boson stars and $1^+$ boson stars
for gravitational coupling $\kappa=0.1$ (left) and $0.2$ (right).
}
\label{QrotPM}
\end{figure}

In Fig.~\ref{QrotPM} we compare the $1^-$ boson stars 
to the $1^+$ boson stars.
In particular, we exhibit the charge $Q$ versus the frequency
$\omega_s$
for gravitational coupling $\kappa=0.1$ and $0.2$.

\begin{figure}[h!]
\parbox{\textwidth}
{\centerline{
\mbox{
\epsfysize=10.0cm
\includegraphics[width=70mm,angle=0,keepaspectratio]{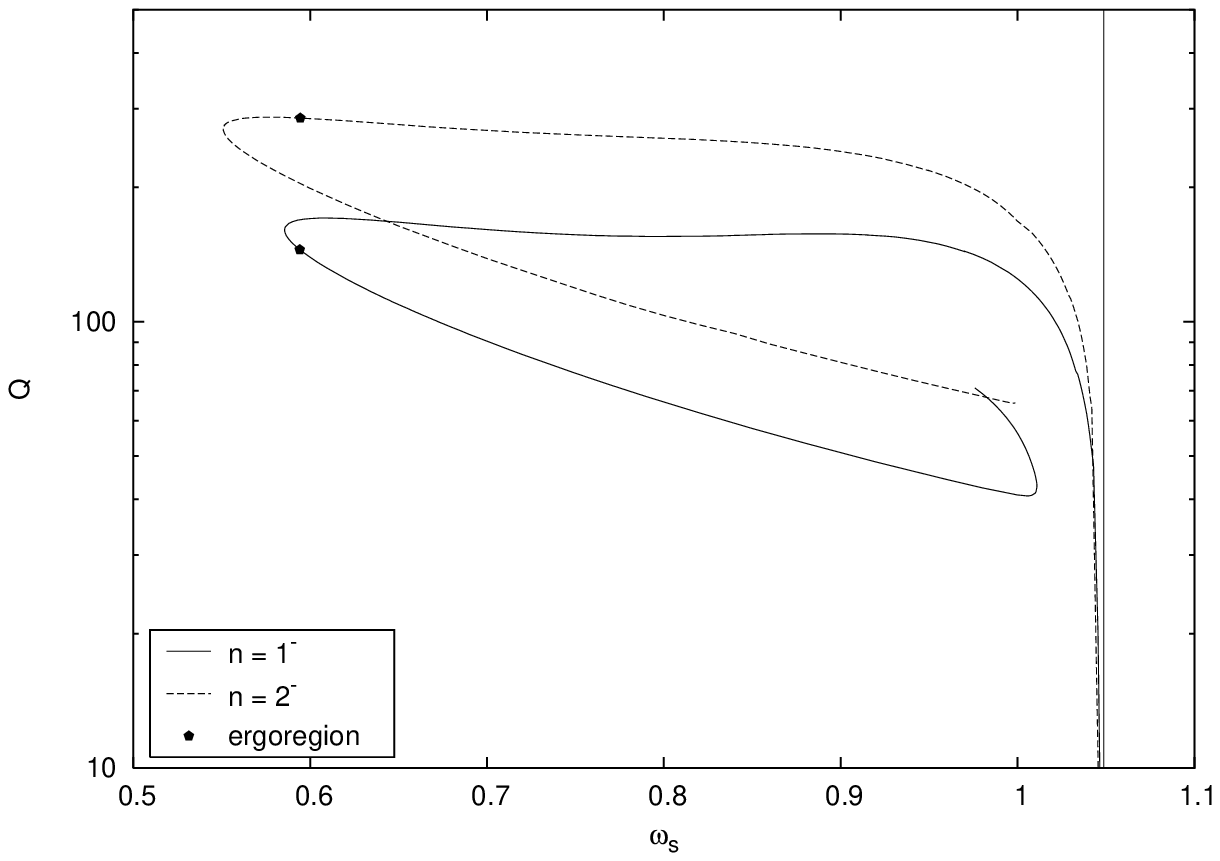}
\includegraphics[width=70mm,angle=0,keepaspectratio]{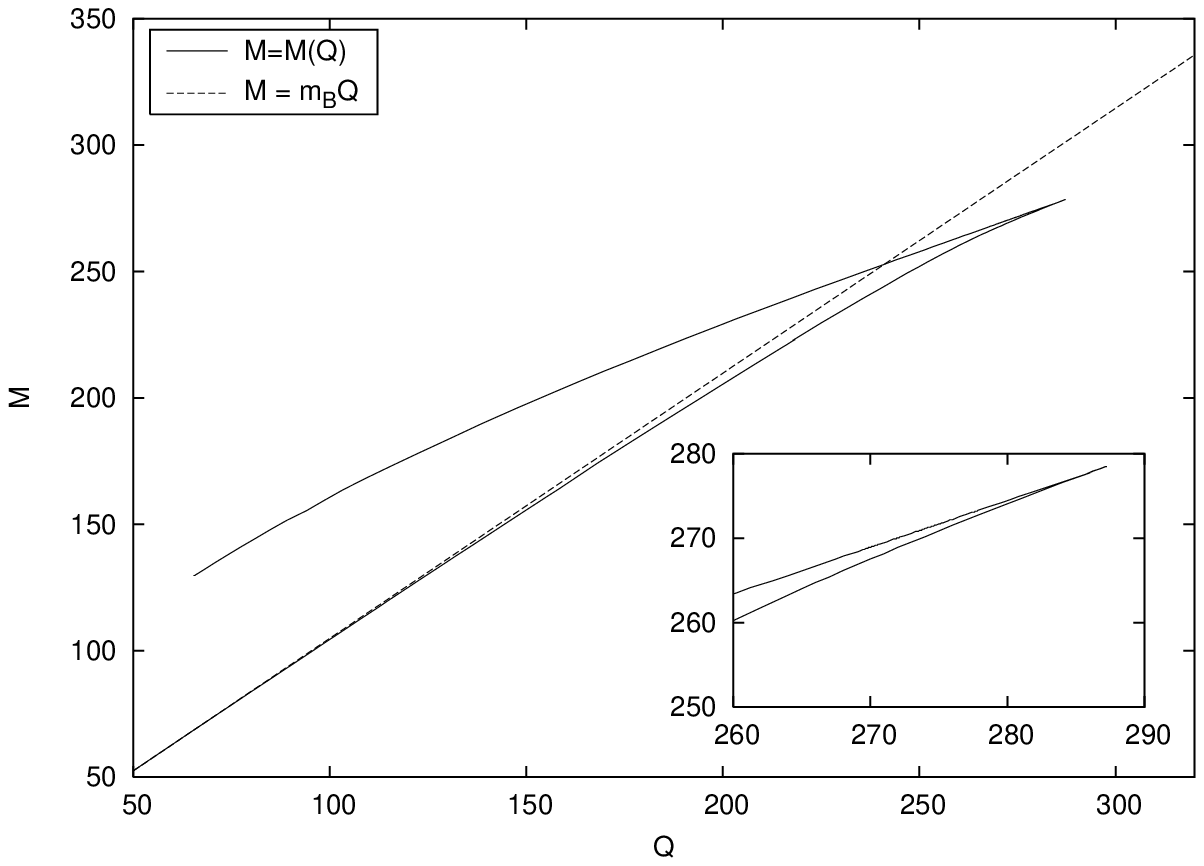}
}}}
\caption{
Left: The charge $Q$ versus the frequency $\omega_s$
for $1^-$ and $2^-$ boson stars
for gravitational coupling $\kappa=0.2$.
The dots indicate the onset of ergoregions.
Right: The mass $M$ versus the charge $Q$
for $2^-$ boson stars
for gravitational coupling $\kappa=0.2$
together with the mass for $Q$ free bosons, $M=m_{\rm B}Q$.
}
\label{Qrotbs-}
\end{figure}

The $n$-dependence of the negative parity boson stars
is demonstrated in Fig.~\ref{Qrotbs-}.
Here the charge $Q$ is exhibited versus the frequency $\omega_s$
for coupling constant $\kappa=0.2$.
The figure also shows the mass $M$ versus the charge $Q$
for $2^-$ boson stars for $\kappa=0.2$, exhibiting
part of the cusp structure.

Considering the spatial structure of the boson stars 
with negative parity,
the scalar field $\f$ here typically gives rise to
a double torus-like energy density $T_{tt}$ 
for the $1^-$ and $2^-$ boson stars on the main branch,
seen already in the negative parity $Q$-balls.
As in flat space, the radii of the double tori of the energy density 
increase when $n$ and thus the angular momentum increases.
In the spiral the spatial structure of solutions
can become more complicated, exhibiting for instance a quadruple
torus-like structure.

\begin{figure}[h!]
\parbox{\textwidth}
{\centerline{
\mbox{
\epsfysize=10.0cm
\includegraphics[width=75mm,angle=0,keepaspectratio]{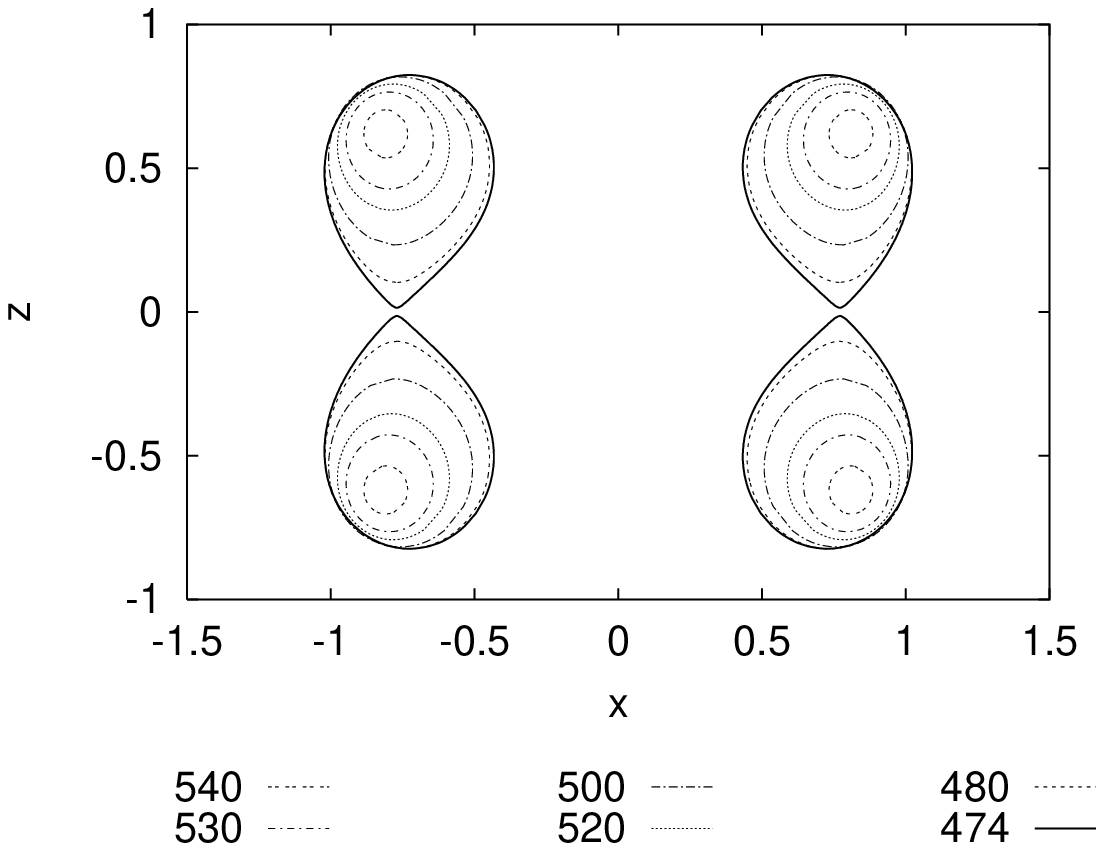}
\epsfysize=10.0cm
\includegraphics[width=75mm,angle=0,keepaspectratio]{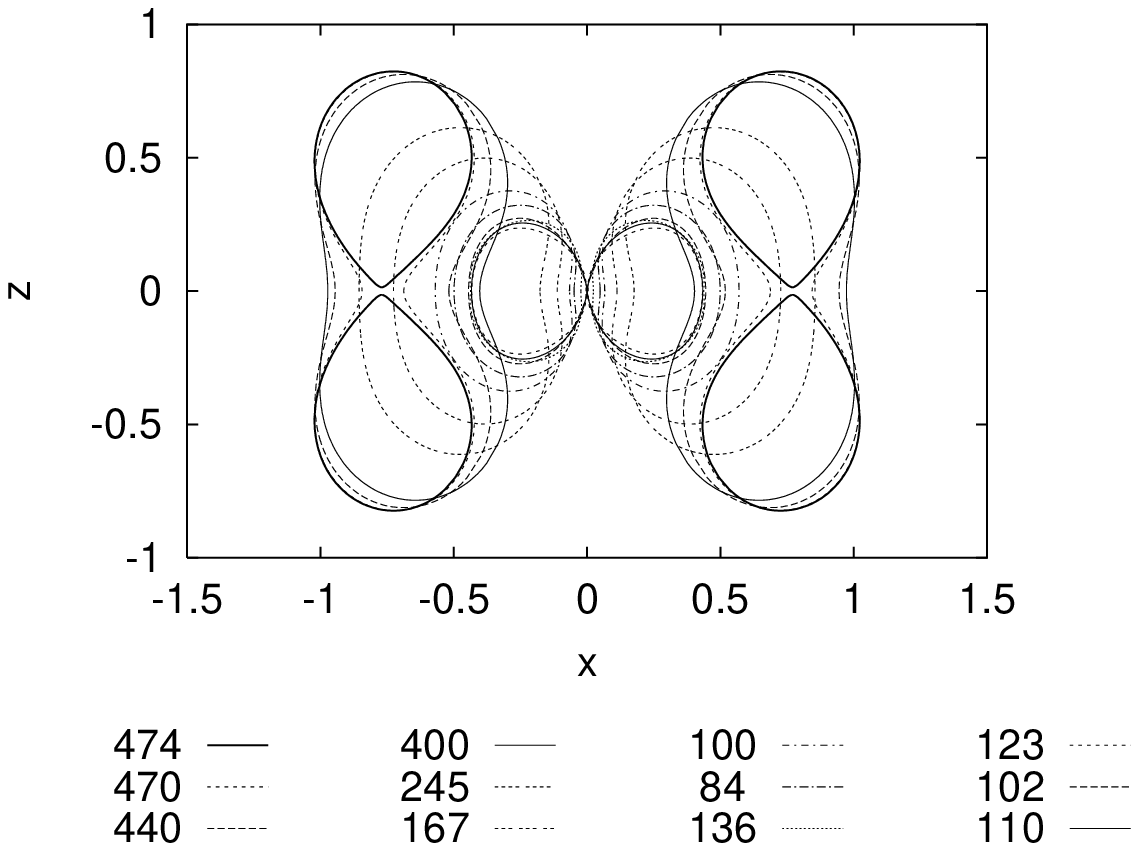}
}}}
{\centerline{
\mbox{
\epsfysize=10.0cm
\includegraphics[width=75mm,angle=0,keepaspectratio]{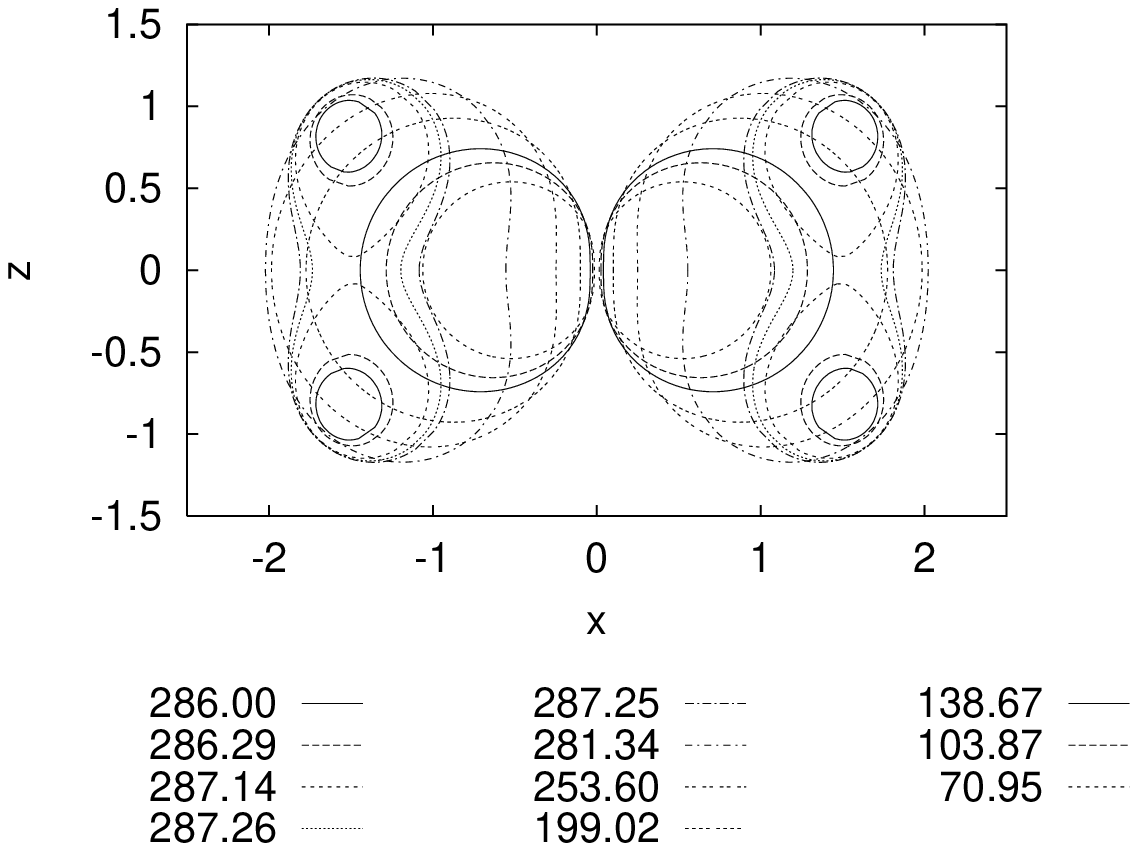}
}}}
\caption{
Location and shape of the axially symmetric ergoregions
of boson stars exhibited in the $xz$-plane:
$1^-$ boson stars at $\kappa=0.1$ (upper row)
and $2^-$ boson stars (lower row)
at $\kappa=0.2$.
The numbers in the legend indicate the respective values of the charge $Q$.
}
\label{ergorot-}
\end{figure}

Addressing the occurrence of ergoregions 
for the above sets of boson star solutions,
we observe that rotating boson stars with negative parity also
possess ergoregions in a large part of their domain of existence.
The onset of the ergoregions for the $1^-$ and $2^-$ 
boson star solutions 
is again indicated by dots in the respective figures.
Again, if the onset occurs on the main branch,
then the solutions 
to the left of the dot and in the spiral do possess an ergoregion,
while the solutions to the right of the dot do not possess an ergoregion.
Likewise, if the onset occurs only in the spiral, then the solutions further
down the spiral all possess ergoregions, while the other solutions do not.
We demonstrate the emergence and structure of these ergoregions
in more detail in Fig.~\ref{ergorot-},
where we exhibit the ergoregions
of $1^-$ boson stars at $\kappa=0.1$
and $2^-$ boson stars at $\kappa=0.2$.

Addressing first the $1^-$ boson stars, we observe, that 
at a critical value of the frequency (where $Q\approx 540$ for $\kappa=0.1$)
the condition (\ref{ergo-c})
is satisfied precisely on two circles,
located symmetrically w.r.t.~the equatorial plane.
As seen in the (upper left) figure,
the circles turn into two torus-like surfaces,
when $Q$ is decreased in the first part of the spiral.
The ergoregion thus consists of two disconnected tori,
after it first arises.

Following the set of solutions further into the spiral,
the two parts of the ergoregion increase in size
until they touch at a critical value of the charge
($Q\approx 447$ for $\kappa=0.1$)
in the equatorial plane, while their centers
shift only little. Beyond this critical value,
the ergoregion then consists of only one connected piece,
a topological torus with an $8$-shaped cross-section.

As seen in the (upper right) figure,
the ergoregion then rapidly changes shape
and assumes a more ellipsoidal cross-section,
when 
the solutions evolve deeper into the spiral.
At the same time, the center of the torus
moves further inwards.
Again the innermost part of the ergosurface gets increasingly
closer to the origin, but never touches it.

The ergoregions of the set of boson stars at different values
of the gravitational coupling constant $\kappa$
exhibit an analogous behaviour.
Likewise, the ergoregions of boson stars at higher $n$ also
exhibit an analogous behaviour 
as seen in the (lower part of the) figure. 

\section{Conclusions}

We have addressed boson stars and their flat space limit, $Q$-balls,
recalling and extending known results for positive parity solutions
and presenting new negative parity solutions.
Our main emphasis has been to study the general pattern
displayed by these regular extended objects, 
to determine their domain of existence, and to investigate their
physical properties.

$Q$-balls and boson stars
exist only in a limited frequency range.
Whereas both mass and charge of $Q$-balls assume a minimal value
at a critical frequency,
from where they rise monotonically towards both smaller and larger
frequencies,
boson stars show a different type of behaviour.
Their mass and charge tend to zero when the maximal frequency is approached,
while for smaller values of the frequency
the charge and mass of boson stars exhibit a spiral-like frequency dependence.
The spirals end in limiting solutions with finite values of 
the mass and charge,
which depend on the gravitational coupling constant $\kappa$.

$Q$-balls and boson stars possess radial excitations, 
which so far have only been studied for spherically symmetric solutions,
and even there a systematic study has not yet been performed. 
Preliminary results indicate, that
radially excited $Q$-balls and boson stars
possess an analogous frequency dependence 
as the corresponding fundamental solutions.

Rotating $Q$-balls and boson stars possess an angular momentum $J$
which is quantized in terms of the charge $Q$, $J=nQ$,
with integer $n$.
The positive parity solutions $n^+$ exhibit a torus-like energy density,
while the negative parity solutions $n^-$ 
exhibit a double torus-like energy density,
with the two tori located symmetrically w.r.t.~the equatorial plane.
%
Clearly, the double torus-like structure of the negative parity solutions
is energetically unfavourable as compared to the single torus-like structure
of the positive parity solutions. 

The classical stability of $Q$-balls and boson stars 
can be analyzed according to catastrophe theory,
implying a change of classical stability at each cusp encountered,
when the mass is considered as a function of the charge.
Both type of solutions then possess classically stable branches,
representing the physically most relevant sets of solutions,
which in the case of boson stars
have potential applications to astrophysics.

Rotating objects may possess an ergoregion.
But for globally regular objects such as boson stars
the presence of an ergoregion would imply an instability,
associated with superradiant scattering
\cite{Cardoso,ergo-paper1,ergo-paper2,ergo-paper3}.
Thus rotating boson stars should become unstable,
when they develop an ergoregion.
Therefore the possible presence of ergoregions 
was put forward by Cardoso et al.~\cite{Cardoso}
as an argument to exclude boson stars and various other black hole doubles
as potential horizonless candidates for compact dark astrophysical objects.

Our analysis has shown that ergoregions can indeed be present
for boson star solutions on the classically stable branch.
But their presence only diminishes the set of boson star solutions
with possible physical relevance
(where the diminishment is greater
for larger gravitational coupling and larger angular momentum),
while there always remains a part of the branch
of classically stable boson star solutions,
not suffering from an ergoregion instability.
Therefore, rotating boson stars cannot yet be excluded 
as potential candidates for compact astrophysical objects.
It remains to be seen, whether boson stars with
appropriate values of the physical parameters
to fit observational data
will or will not suffer from such an ergoregion instability.

\begin{acknowledgments}
We gratefully acknowledge V.~Cardoso for drawing
our attention to the ergoregions of boson stars.
B.K. gratefully acknowledges support by the German Aerospace Center
and by the DFG.
\end{acknowledgments}


\end{document}